\documentclass{aa} 

\usepackage{natbib}
\usepackage{pdflscape}
\usepackage{etoolbox}
\usepackage{amsmath}
\usepackage{xspace}
\usepackage[utf8]{inputenc}
\usepackage{caption}
\usepackage{subcaption}
\usepackage[varg]{txfonts}
\usepackage{multirow}
\usepackage[dvipsnames]{xcolor}
\usepackage{graphicx}
\usepackage[breaklinks=true,colorlinks,citecolor=blue,urlcolor=blue]{hyperref}
\usepackage[round-mode=places,round-precision=2,group-separator={,},output-decimal-marker={.}]{siunitx}

\bibpunct{(}{)}{;}{a}{}{,}
\graphicspath{{./}{figures/}}

\makeatletter
\let\clandscape=\landscape

\patchcmd{\clandscape}{\PLS@Rotate{90}}{\PLS@Rotate{-90}}{}{}
\makeatother

\newcommand{\msun}{\ensuremath{\mathrm{\,M_\odot}}} 
\newcommand{\rsun}{\ensuremath{\mathrm{\,R_\odot}}} 
\newcommand{\mdotsun}{\ensuremath{\mathrm{\,M_{\odot}\,yr^{-1}}}} 

\newcommand{\mdot}{\ensuremath{\dot{M}}} 

\newcommand{\mesa}{{\textsc{mesa}}\xspace} 

\begin{document} 

\title{Formation of twin compact stars in low-mass X-ray binaries}

\subtitle{Implications for eccentric and isolated millisecond pulsar populations}

\author{ S. Chanlaridis\inst{1,2} 
        \and D. Ohse\inst{3} 
        \and D. E. Alvarez-Castillo\inst{4,5,6,7,8}
        \and J. Antoniadis\inst{1,9}
        \and D. Blaschke\inst{10,11,12}  
        \and \\ V. Danchev\inst{13,14}
        \and N. Langer\inst{3,9}
        \and D. Misra \inst{15}
        }

   \institute{Institute of Astrophysics, Foundation for Research \& Technology -- Hellas (FORTH), GR-70013 Heraklion, Greece \\
   \email{schanlaridis@physics.uoc.gr}
            \and
            Department of Physics, University of Crete, University Campus, GR-70013 Heraklion, Greece 
            \and
            Argelander-Institut f\"{u}r Astronomie, Auf dem H\"{u}gel 71, DE-53121 Bonn, Germany 
            \and
            Institute of Nuclear Physics, Polish Academy of Sciences, Radzikowskiego 152, 31-342 Cracow, Poland
            \and
            Incubator of Scientific Excellence - Centre for Simulations of Superdense Fluids, Max Born place 9, 50-204 Wroclaw, Poland
            \and
            Facultad de Ciencias Físico Matemáticas, U.A.N.L., Av. Universidad S/N, C.U., 66455 San Nicolás de los Garza, N.L., Mexico
            \and
            Helmholtz Institute Mainz, 55099 Mainz, Germany
            \and
            GSI Helmholtzzentrum für Schwerionenforschung GmbH, 64291 Darmstadt, Germany
            \and
            Max-Planck-Institut f\"{u}r Radioastronomie, Auf dem H\"{u}gel 69, DE-53121 Bonn, Germany
            \and
            Institute of Theoretical Physics, University of Wroclaw, Max Born place 9, 50-204 Wroclaw, Poland
            \and
            Center for Advanced Systems Understanding (CASUS),
            Untermarkt 20, DE-02826 G\"orlitz, Germany
            \and
            Helmholtz Zentrum Dresden Rossendorf (HZDR),
            Bautzener Landstra{\ss}e 400, DE-01328 Dresden, Germany
            \and
            Department of Physics, Sofia University St. Kliment Ohridski, 5 James Bourchier Blvd, 1164 Sofia, Bulgaria 
             \and
             EnduroSat, 1A Flora Street, 1404, Manastirski Livadi, Sofia, Bulgaria
            \and
             Institutt for Fysikk, Norwegian University of Science and Technology, 7491 Trondheim, Norway 
            \\
             }

   \date{Received; accepted}

\abstract
{Millisecond pulsars (MSPs) are laboratories for stellar evolution, strong gravity, and ultra-dense matter. Although MSPs are thought to originate in low-mass X-ray binaries (LMXBs), approximately 27\% do not have a binary companion, and others are found in systems with large orbital eccentricities. Understanding how these systems form may provide insight into the internal properties of neutron stars (NSs).}
{We studied the formation of a twin compact star through  rapid first-order phase transitions in NS cores due to mass accretion in LMXBs. We investigated whether this mechanism, possibly coupled with secondary kick mechanisms such as neutrino or electromagnetic rocket effects, leaves an observable long-lasting imprint on the orbit.}
{We simulated mass accretion in LMXBs consisting of a NS and a low-mass main-sequence companion and followed the evolution of the NS mass, radius, and spin until a strong phase transition is triggered. For the internal NS structure, we assumed a multi-polytrope equation of state  that allows a sharp phase transition from hadronic to quark matter and satisfies observational constraints.}  
{We find that in compact binary systems with relatively short pre-Roche lobe overflow orbital periods, an accretion-induced phase transition can occur during the LMXB phase. In contrast, in binary systems with wider orbits, this transition can take place during the spin-down phase, leading to the formation of an eccentric binary MSP. If the transition is accompanied by a secondary kick with a magnitude $> 20$\,km\,s$^{-1}$, then the binary has a high probability of being disrupted, thereby forming an isolated MSP, or being reconfigured into an ultra-wide orbit.} 
{Our findings suggest that accretion in LMXBs provides a viable path for the formation of twin compact stars that could leave a long-lived and thus observable imprint on the orbit. The eccentricity distribution of binary MSPs with long orbital periods  ($>50\,$d) can provide stringent constraints on first-order phase transitions in dense nuclear matter.} 


\authorrunning {}
\titlerunning {Twin compact stars in LMXBs}
\maketitle

\section{Introduction} \label{sec:intro}
Millisecond pulsars (MSPs) are widely considered to be old, spun-up neutron stars (NSs) characterized by high rotational frequencies, $\mathcal{O}(10^2\,\rm Hz)$, and weak surface magnetic fields, $\mathcal{O}(10^8\,\rm G)$. The formation of MSPs can be traced back to a recycling
phase in a low-mass X-ray binary \citep[LMXB; see][]{Tauris:2023nmj}. 
During the recycling process,  mass and angular momentum are accreted onto the NS from the companion star when the latter fills its Roche lobe \citep[e.g.,][and references therein]{Bhattacharya:1991pre, Tauris:aap1999, Tauris:2023nmj}. During mass transfer,  tidal interactions circularize the orbit on a timescale that is much shorter $(\sim 10^4\,\rm yr)$ than the  mass accretion phase. Therefore, MSPs are expected to populate binary systems with very small eccentricities \citep{Phinney:1992, Verbunt:aa1995}.

Although the vast majority of MSPs are indeed situated in systems with highly circularized orbits, a considerable fraction  appear to be solitary \citep[$\sim 27$\% of known MSPs have spin periods $\leq 30$\,ms; see the ATNF pulsar catalog;][]{Manchester_2005}\footnote{\url{https://www.atnf.csiro.au/research/pulsar/psrcat/}\\ accessed on 29 May 2024}. These isolated MSPs (iMSPs) raise questions about the recycling process, particularly about possible mechanisms that could eject, consume, or disrupt the MSP companion stars \citep{1987Natur.329..312V, 2019PASA...36....5S, 2019JApA...40...32N, 2020A&A...633A..45J, antoniadis:2021}. 

In addition, recent pulsar surveys have revealed the existence of binary MSPs with eccentricities of up to $e \simeq 1$. With
the exception of MSPs in globular clusters and PSR\,J1903+0327 \citep{Champion:sci2008}, which is a Galactic field MSP with a
main-sequence companion and an eccentricity of $e\simeq0.4$, 
 eccentric MSPs (eMSPs) appear to have orbital periods between 20 and 50\,d and eccentricities of $\mathcal{O}(0.1)$, 
 \citep{Deneva:apj2013, Barr:mnras2013, Knispel:apj2015, Camilo:apj2015, antoniadis:apjl14, Antoniadis:2016bnj,  Stovall:apj2019}. 
Although chaotic gravitational interactions, such as the ejection of the least massive member of a hierarchical triple system
progenitor, have the potential to explain the formation of eMSPs in clusters \citep[e.g.,][]{2011MNRAS.412.2763F, 
2011ApJ...734...55P},   they are unlikely to produce systems that have similar binary parameters \citep[see][]
{Deneva:apj2013, Barr:mnras2013, Knispel:apj2015, Camilo:apj2015, Octau:aap2018}, such as those found in the Galactic field. 
However, several attempts have been made to investigate alternative formation channels that can accommodate the observed properties of eMSPs \citep[e.g., see ][for an overview of various eMSP formation scenarios]{Freire:mnras14, antoniadis:apjl14, Jiang:apj15, Jiang:raa2021, ginzburg:mnras21}. 

The formation of both iMSPs and eMSPs could be related to abrupt changes in the interior of the NS \citep{Freire:mnras14, Jiang:apj15, Alvarez-Castillo:2019apz}. For instance, the dependence of the NS structure on the equation of state (EoS) highlights the possibility of exotic phases and phase transitions in their interiors. 
\cite{Jiang:apj15} find that eccentricities of $\sim 0.11 - 0.15$ will be induced by instantaneous gravitational mass loss of the NS due to a phase transition in its interior, even without invoking an explicit kick mechanism.
However, for the abovementioned scenarios, it can be troublesome to explain the ``instantaneity'' of the transition when the NS spin is included. As pointed out by \cite{Glendenning:1997fy} in the context of their scenario for an accretion-induced phase transition to a hybrid star branch that is connected with the NS one, the reconfiguration of the mass distribution in the NS interior is accompanied by a change in the moment of inertia. This would entail a ``pirouette effect'' (a spin-up), which in turn would inhibit the completion of the transition due to the sensitive dependence of the density profile inside the star on its spin state caused by centrifugal forces. \cite{Glendenning:1997fy} obtain in their scenario a timescale of $10^5$ years for the completion of the phase transition, which is in stark contrast with the requirement that the mass defect occur on a timescale much shorter than the orbital period.   

A phase transition scenario that circumvents this problem assumes the existence of a third family of compact stars \citep[henceforth ``twin stars''; see][]{Gerlach:1968zz}
beyond the conventional  white dwarf (WD) and NS classifications. 
Such a third family could emerge as a result of a sufficiently strong phase transition in the NS core, leading to the formation of hybrid stars with an outer core of hadronic matter and an inner core of de-confined quark matter \citep[e.g.,][]
{Schertler:2000xq,Benic:2014jia, Alvarez-Castillo:2019apz, 2021AN....342..234A}. 
Given that baryon number and angular momenta can be conserved simultaneously during this process, the transition can occur almost instantaneously, on a dynamic timescale, which is insignificant compared to the orbital period.

Within the context of MSP formation scenarios, such an instantaneous transition to the third-family branch could be triggered by mass accretion during, or after, the LMXB phase \citep[see, e.g.,][]
{Zdunik:2005kh,Bejger:2016emu,Alvarez-Castillo:2019apz}. This transition could in turn induce instantaneous mass loss and/or a kick that can cause significant changes to the orbital dynamics \citep{Jiang:apj15}. 
More specifically, the sudden decrease in gravitational mass due to quark de-confinement in the core \citep[owing to the 
``catastrophic rearrangement'' and strong compactification of matter when transitioning to the third-family branch, which increases the gravitational binding energy; see][]{Mishustin:2002xe} may significantly increase eccentricity or even disrupt the binary \citep[see also][]{Jiang:raa2021}. The main aim of this work is to explore this possibility in more detail. 

The investigation of such scenarios is important for advancing our understanding of dense matter under extreme conditions.  Recent advances in observational techniques, such as gravitational wave (GW) detections from NS mergers \citep[e.g.,][]{PhysRevLett.119.161101} and high-precision pulsar timing
\citep[e.g.,][]{Miller:2019cac, Riley:2019yda,Miller:2021qha,Bogdanov:2021yip}, offer unprecedented opportunities to test and refine theoretical models that predict phase transitions \citep[e.g.,][]{Bauswein:2022vtq}.
For instance, GWs from unstable fundamental $f$-mode oscillations can reveal the internal stellar
structure and matter distribution characteristic of a first-order phase transition, as shown in \cite{2023arXiv230908775K},
where estimations for the same class of model used in this work are presented. 

\cite{2023arXiv231115992K} simulated NS evolution with an accretion-induced phase transition in a neutrino radiation hydrodynamics treatment,
implementing a three-flavor Boltzmann neutrino transport and a microscopic quark-hadron hybrid model EoS. 
This approach results in the production of gravitational radiation from $f$-mode excitation and the associated neutrino emission. 
The phase transition triggers an expansion of the accreted envelope with supersonic velocities resulting in a mass loss of $\sim 10^{-3}\msun$. 
It is worth noting that finite temperatures are considered in this approach, which correspond to an entropy per baryon of $0.1-0.5$. 
Since finite temperatures soften the quark matter EoS, the onset densities for de-confinement are lowered and the latent heat is increased so that twin stars of identical mass but different radii and compactness \citep{Gerlach:1968zz, Schertler:2000xq} become possible even when at the vanishing temperature no gravitational instability is associated with the de-confinement transition \citep{Hempel:2015vlg,Carlomagno:2023nrc}. 
The associated mass defect of such an accretion-induced thermal star quake can be as much as $3\%$ of the solar mass \citep{Carlomagno:2024vvr}.
In fact, with the precision of the next generation of GW detectors, we expect to be able to study these scenarios. 
In general, the timely synergy between theoretical developments and observational capabilities places us at the forefront of uncovering the complex NS physics, where phase transitions emerge as a major component \citep{Bauswein:2022vtq}.

In this work we combined detailed binary evolution calculations with a simple analytic model for the response of the NS spin and radius to accretion to investigate whether twin stars can form in LMXBs. We also investigated whether such accretion-induced phase transitions could help explain the observed eMSP and iMSP populations. The text is organized as follows: In Sect.~\ref{sec:methods} we describe the methodology and physical assumptions. We present the simulation results in Sect.~\ref{sec:results} and conclude with a summary and discussion on the implications of our calculations in Sect.~\ref{sec:summary}.

\section{Methodology} \label{sec:methods}

\subsection{Initial setup and numerical calculations}\label{sec:setup}
Our main goal was to investigate whether first-order phase transitions can be triggered in LMXBs and recycled MSPs under realistic mass accretion and spin evolution scenarios. To this end, our working setup was divided into three distinct components:

\begin{enumerate}
    \item An EoS for dense nuclear matter that allows for a first-order phase transition paired with a mass-radius relation for spinning NSs; 
    \item A stellar evolution model for the binary system, providing the mass-accretion profile as a function of time for the NS; 
    \item An accretion model that allows us to calculate the spin, mass, and radius evolution of the NS during and after the LMXB phase.
\end{enumerate}
They are described in detail in the remainder of the section.

\subsubsection{Equation of state and mass-radius relation}\label{sec:eos}
For the EoS, we selected a multi-polytrope model suitable for describing both nuclear and quark matter within NSs. This EoS was chosen for its ability to (a) account for a strong first-order phase transition from hadronic to quark matter above $1.4\msun$ (depending on the angular momentum) and (b) satisfy current constraints derived from multi-messenger astronomy \citep{antoniadis:2013sci, LIGOScientific:2018cki, Miller:2019cac, Riley:2019yda, 2020NatAs...4...72C}.

The outer layers, characterized by densities $n \leq 0.5 n_0$, where $n_0$ denotes the nuclear saturation density, are modeled using the \texttt{BPS} EoS \citep{Baym:1971apj}. An intermediate density regime with $0.5 n_0 < n \leq 1.1 n_0$ is assumed to consist of homogeneous matter in $\beta$-equilibrium. For regions with densities $n > 1.1 n_0$, we distinguished four regimes, each described by a polytropic EoS of the form
\begin{equation}
    P(n) = \kappa_i \left( \frac{n}{n_0} \right)^{\Gamma_i},
    \label{eq:acb5_eos}
\end{equation}
henceforth referred to as the \texttt{ACB5} EoS as in \cite{Paschalidis:2018prd}.

The first polytrope, $(i=1)$, is a fit to the stiffest EoS version provided in \cite{Hebeler:2013apj}. The second polytrope, $(i=2)$, describes the first-order phase transition. The polytropic index $\Gamma$ allows us to distinguish between a sharp transition (Maxwell construction, $\Gamma = 0$) and a non-sharp transition (Gibbs construction, $\Gamma \neq 0$). Since we required a strong transition to produce a large jump in energy density, the second polytrope was defined as a region of constant pressure $P_c = \kappa_2$, where $\Gamma_2 = 0$ is imposed by the Maxwell construction. The remaining two polytropes $(i = 3,4)$ describe regions with densities that exceed the critical value for a first-order transition and correspond to stiff quark matter that should be able to support a NS with masses up to $\sim 2.0\msun$ \citep[e.g.,][]{2020NatAs...4...72C}.
The different density regimes are thermodynamically joined in a consistent manner. 

To infer the corresponding mass-radius relation for rigidly spinning NSs, we used the numerical integrator described in \cite{Alvarez-Castillo:2019apz}. 
In this code, the integration of the structure equations proceeds radially outward from the center of the NS, where the density is highest, to its surface, where the pressure falls to zero, thus determining the total mass and radius for various central densities and angular momenta, $J$.

\begin{figure*}
    \centering
    \includegraphics[width=\textwidth]{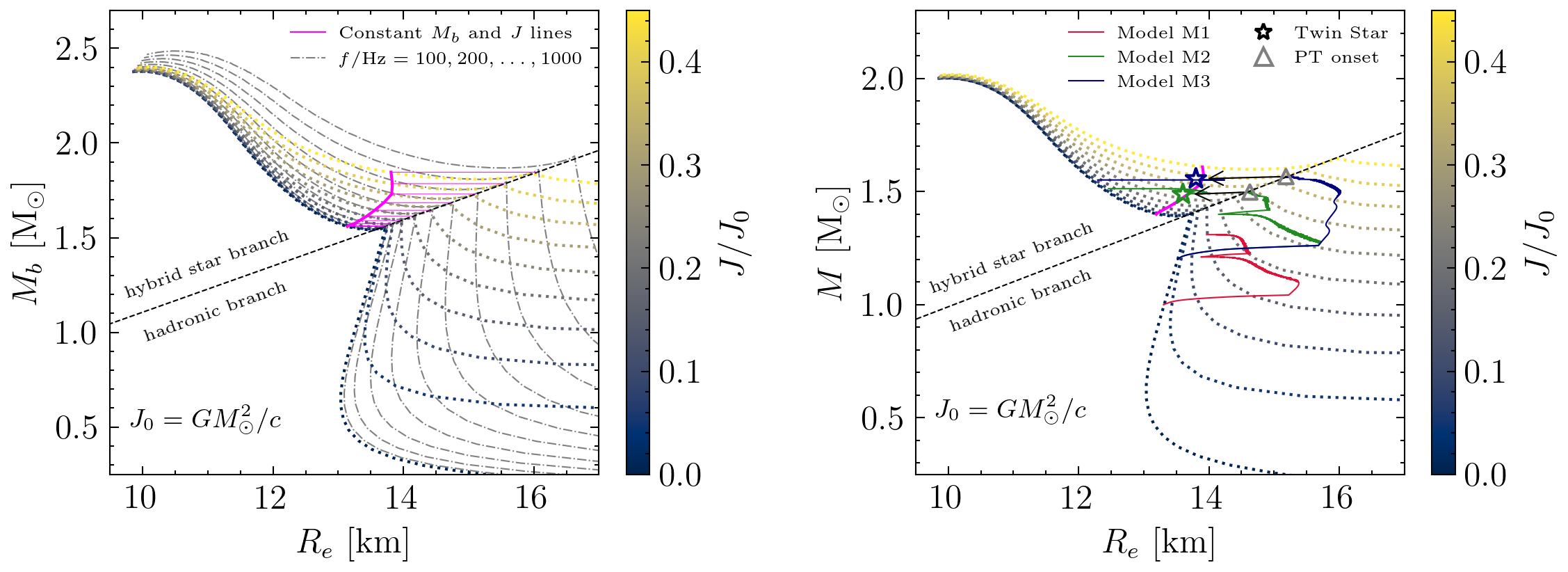}
    \caption{\textit{Left Panel}: Baryonic mass vs. equatorial radius. Dotted and dot-dashed curves indicate lines of constant angular momentum and constant frequency, respectively. The dashed black line highlights the points of maximum stability in the hadronic branches, signaling the onset of a phase transition (PT). Thin, horizontal magenta lines show trajectories where angular momentum ($J$) and baryonic mass ($M_b$) are conserved. The thick magenta curve connects points on the hybrid star branches that can be reached through a collapse conserving both $J$ and $M_b$.
    \textit{Right Panel}: Similar to the left panel, but with the y-axis representing gravitational mass. Star markers denote the endpoint trajectories for direct (\textsc{m2}) and delayed (\textsc{m3}) collapse models. The arrows are inclined because gravitational mass, unlike baryonic mass, is not conserved during the PT.}
    \label{fig:mvsr}
\end{figure*}

The resulting mass-radius relations are illustrated in Fig.~\ref{fig:mvsr} where the baryonic mass (left) and gravitational mass (right) mass are plotted against the equatorial radius of the NS. 
The lower right section of each panel depicts the mass-radius relation for NSs within the hadronic branch, corresponding to the first \texttt{ACB5} polytrope and indicating a composition of purely hadronic matter. Moving toward the upper left segment of each panel, we shift into the twin (hybrid) star branch, representing NSs that have undergone a phase transition to include a quark matter core, described by the \texttt{ACB5}'s third and fourth polytropes. This transition is shown by a dashed black curve, representing the second polytrope of the \texttt{ACB5} EoS and indicating the highest mass that a stable NS composed entirely of hadronic matter can sustain. Beyond this threshold, additional mass triggers a phase transition to quark matter, moving the star to the hybrid branch.

The role of spin and angular momentum in dictating the precise moment of transition is crucial; higher angular momentum allows a NS to support more mass in the hadronic phase. This effect is illustrated by the dotted curves, which denote varying levels of angular momenta, normalized by $J_0=G\msun^2/c$. Here, the innermost (dark blue) dotted line represents a static NS, and the outermost (light yellow) line corresponds to $J/J_0 = 0.45$. Beyond this point, the hadronic branch extends past the shedding limit, where equatorial matter attains Keplerian escape velocity and is ejected.

For the calculations that follow, we assumed that a NS reaching the transition line undergoes a phase transition on the dynamic timescale, under conservation of the angular momentum and baryonic mass. 
The transition moves the NS from the maximal stability threshold on the hadronic branch (dashed line), to a corresponding position (in terms of angular momentum and baryonic mass) on the third family. This principle is visually represented by the magenta curve, linking points on the hybrid branch that could be attained through the transition and defining the birth radius of the twin star. 

\subsubsection{Mass transfer and binary evolution models} \label{sec:input_params}
To obtain realistic mass-transfer models, we performed detailed numerical binary calculations using the implicit one-dimensional code Modules for Experiments in Stellar Astrophysics \citep[\mesa\,v12778;][]{Paxton:2010ji,Paxton:2013pj,Paxton:2015jva,Paxton:2017eie}. Our binary systems consisted of a zero-age main-sequence (ZAMS) donor with an initial metallicity $Z=0.02$ and an NS accretor, which was treated as a point mass.
We calculated three models chosen as representative of three distinct cases: an LMXB in which a phase transition never occurs (\textsc{m1}), one in which the transition occurs during the mass-transfer phase (\textsc{m2}), and a model for which the transition occurs after the accretion phase, due to the spin-down of the NS (\textsc{m3}). The initial binary parameters for each model are summarized in Table~\ref{tab:config}.

\begin{table}[t]
    \centering
    \caption{Initial binary parameters for our stellar evolution models.}
    \begin{tabular}{cccc}
         \hline \hline \\
         Model & $M_\mathrm{don}\,[\msun]$ & $M_\mathrm{ns}\,[\msun]$ & $P_\mathrm{orb}\,[\mathrm{days}]$  \\
         \hline \\
        \textsc{m1} & $1.0$ & $1.0$ & $8$ \\\\
        \textsc{m2} & $1.0$ & $1.2$ & $8$ \\\\
        \textsc{m3} & $1.0$ & $1.2$ & $22.627$ \\\\
        \hline
    \end{tabular}
    \label{tab:config}
\end{table}

For the donor-star evolution, we used the type-2 OPAL Rosseland mean opacity tables \citep{opal_opacities}. Convection was modeled using the modified mixing length theory prescription of \cite{Henyey:apj1965} with a mixing length parameter of $\alpha_{\rm ML} = 1.8$. In our models, we enabled convective premixing while trying to avoid increases in the abundance of species that were burned, where possible. Stability against convection was determined according to the Ledoux criterion \citep[][]{Ledoux:apj1947}. Furthermore, we used semi-convection, which we treated as a diffusive process, with an efficiency parameter of $\alpha_{\rm SEM} = 1.0$ \citep[][]{Langer:aap1991}. Lastly, we accounted for the overshooting of convective material beyond the convective layers by adopting an exponential overshooting efficiency of $f_{\rm ov, core} = 0.016$ and $f_{\rm ov, env} = 0.0174$ in the core and envelope, respectively. 

Rotational mixing was taken into consideration, including the effects of Eddington–Sweet circulations, secular and dynamical instability, and the Goldreich–Schubert–Fricke instability \citep[see][for details]{Heger:apj2000}. The contribution of these instabilities to the total diffusion coefficient was reduced by a mixing efficiency factor, $f_c = 1/30$ \citep[][]{1992A&A...253..173C, Heger:apj2000}. Moreover, we employed the Spruit-Tayler dynamo to compute the internal magnetic field strength and the corresponding transport of angular momentum, as described in \cite[][]{Spruit:aap2002, Heger:apj2005}.

To calculate the mass loss rate due to stellar wind, we implemented the cool red giant branch wind scheme from \cite{Reimers:bk1975}:
\begin{equation}
    \label{eq:reimers_mdot}
    \dot{M}_{1, \text{wind}} = -4 \times 10^{-13} \eta \left(\frac{R_1}{R_\odot}\right) \left(\frac{L_1}{L_\odot}\right)\left(\frac{M_\odot}{M_1}\right)\quad [M_\odot\,\text{yr}^{-1}],
\end{equation}
using a scaling factor of $\eta = 0.1$. In cases of hydrogen-poor donor stars whose hydrogen mantles have been stripped (surface hydrogen mass fraction $X_{\rm S} < 0.4$), we applied the prescription of \cite{Nugis:aap2000} with a scaling factor of $\eta = 1.0$. This factor is used in \mesa under the \texttt{Dutch} hot wind scheme \citep[][]{Glebbeek:aap2009}.

We assumed that our binary models were initially tidally synchronized to the orbit at the beginning of the evolution (ZAMS stage) and that the initial eccentricity was negligible. These assumptions are justified as tidal forces would circularize (and synchronize) the orbit on a much shorter timescale compared to the main-sequence lifetime of a low-mass donor star \citep[e.g.,][]{Verbunt:aa1995}.

To compute the mass transfer rates during Roche lobe overflow (RLOF), we used the prescription suggested by \cite{Kolb:aap1990}. In addition, an isotropic reemission scenario of mass transfer was adopted \citep[see][for a review]{Tauris:bk2006}. In this scenario, we assumed that mass flows conservatively from the donor to the NS accretor via RLOF, and a fraction of the transferred material, $\beta$, is lost (reemitted) from the vicinity of the NS as an isotropic fast wind, carrying away the specific angular momentum of the NS. Hence, the NS's efficiency of accretion, $\epsilon$, is defined as
\begin{equation}
    \label{eq:accretion_efficiency}
    \epsilon = 1 - \alpha - \beta - \delta,
\end{equation}
where $\alpha$ is the fraction of mass lost directly from the donor via winds, and $\delta$ is the fraction of mass lost from a circumbinary coplanar toroid. Here, we omitted any losses from winds or circumbinary toroids ($\alpha = \delta = 0$) and assumed that 50\% of the transferred mass is reemitted ($\beta = 0.5$) resulting in an accretion efficiency of $\epsilon = 0.5$.
Furthermore, we assumed that the accretion rate of the NS is limited by the Eddington mass accretion rate, which for accreted material composed of pure ionized hydrogen takes the form
\begin{equation}
    \label{eq:edd_limit}
    \dot{M}_{\text{Edd}} = 1.5 \times 10^{-8} \left(\frac{M_{2,i}}{1.3 M_\odot}\right)\quad[M_\odot\,\text{yr}^{-1}],
\end{equation}
where $M_{2,i}$ is the initial mass of the accretor \citep{2020A&A...642A.174M}. Here, the mass accreted would not change the Eddington limit significantly, and thus we fixed the Eddington limit to $\dot{M}_{\text{Edd}} = 1.5 \times 10^{-8}\,M_\odot\,\text{yr}^{-1}$. Excess material, exceeding the Eddington accretion rate, was assumed to carry the specific angular momentum of the NS.

For orbital angular momentum losses, we considered  mechanisms such as GW radiation, mass lost from the system, spin-orbit coupling due to tidal effects, and magnetic braking for donors with convective envelopes.
The orbital angular momentum loss due to GW radiation were calculated using the formula
\begin{equation}
    \label{eq:jdot_gw}
    \frac{dJ_{\text{gw}}}{dt} = - \frac{32}{5}\frac{G^{7/2}}{c^5}\frac{M_1^2 M_2^2 (M_1 + M_2)^{1/2}}{\alpha^{7/2}},
\end{equation}
where $G$ is the gravitational constant, $c$ is the speed of light in vacuum, $\alpha$ is the binary separation, and $M_1, M_2$ are the masses of the donor and the accretor, respectively \citep{Tauris:bk2006}.
Angular momentum loss due to magnetic braking was computed following the prescription from \cite{Rappaport:apj1983}:
\begin{equation}
    \label{eq:jdot_mb}
    \frac{dJ_{\text{mb}}}{dt} = -3.8 \times 10^{-30} M_1 R_\odot^4 \left(\frac{R_1}{R_\odot}\right)^{\gamma}\Omega_1^3\quad\text{[dyn cm]},
\end{equation}
where $M_1, R_1$, and $\Omega_1$ are respectively the mass, radius, and angular velocity of the donor star. Here, the magnetic braking index was fixed to  $\gamma = 4$.

We let our models evolve until the donor star forms a WD or the number of computational steps needed for convergence exceeds a limit of $3\times 10^5$. In all cases, the spatial and temporal resolution were set to their default values. Nonetheless, to assess whether rapid torque fluctuations (see Sect.~\ref{sec:torque_fluctuations}) during the equilibrium phase are physical or numerical in nature, and whether they impact our conclusions, we performed a time-resolution sensitivity study, by running extra \mesa tracks. To this end, we varied the time resolution during the mass-transfer phase of our models across a range of values, from $8\times10^{-5}$ to $8\times10^{-3}$, to test the sensitivity of the spin evolution and check the robustness of our conclusions.

\subsubsection{Evolution of NS spin and structure}
To simulate the response of the interior of the NS to mass accretion, we adopted
the accretion model of \cite{Tauris:sc2012}, which allowed us to calculate the accretion torque as a function of time, $N(t)$. We summarize this model here for the sake of completeness.  

The torque model is based on two primary assumptions: (a) spherical accretion, which simplifies the inflow dynamics to a scenario where the flow of ionized gas is predominantly governed by gravity until it reaches the magnetosphere,
and (b) the magnetic dipole assumption, which posits that the NS surface magnetic field can be approximated as a dipole.
As matter falls on the surface of the NS, the ram pressure of the infalling material is counteracted by the magnetic pressure \citep[e.g.,][]{1996ApJ...457L..31C}. Assuming a spherical accretion scenario, these two pressures can be equated to determine the so-called magnetospheric radius, 
\begin{multline}
    \label{eq:rmag}
    R_\mathrm{mag} \approx 7.8 \left(\frac{B}{10^8\,\mathrm{G}}\right)^{4/7} \left(\frac{R_\mathrm{ns}}{10\,\mathrm{km}}\right)^{12/7} \\ \times \left(\frac{M_\mathrm{ns}}{1.4\msun}\right)^{-1/7} \left(\frac{\mdot}{\mdot_\mathrm{Edd}}\right)^{-2/7}\,[\mathrm{km}],
\end{multline}
inside of which the plasma flow is dictated by the magnetic field corotating with the NS \citep{10.1111/j.1365-2966.2005.09167.x}. This radius provides an estimate of the location of the inner edge of the accretion disk. The outer boundary of the magnetosphere is indicated by the light-cylinder radius:
\begin{equation}
    \label{eq:rlc}
    R_\mathrm{lc} = \frac{cP_\mathrm{spin}}{2\pi},
\end{equation}
which signifies the location where the corotation velocity equals the speed of light \citep[e.g.,][and references therein]{2004ApJ...606..436R, Tauris:sc2012, 2017ApJ...835....4B}. 

Another important length scale is the corotation radius,
\begin{equation}
    \label{eq:rcor}
    R_\mathrm{co} = (G M_\mathrm{ns})^{1/3}\left(\frac{P_\mathrm{spin}}{2\pi}\right)^{2/3},
\end{equation}
which is defined as the radial distance at which material corotating with the NS attains the  
Keplerian angular frequency \citep[e.g.,][]{1996ApJ...457L..31C, 2017ApJ...835....4B, 
Bhattacharyya:mdpi23}. If $R_\mathrm{mag} < R_\mathrm{co}$, the accreted material follows the magnetic 
field lines and is deposited on the surface, carrying angular momentum and exerting a positive torque 
that tends to spin up the NS. However, if $R_\mathrm{mag} > R_\mathrm{co}$ a centrifugal barrier 
arises, inhibiting the accretion flow as the material accelerates to super-Keplerian velocities 
\citep[e.g.,][and references therein]{2017ApJ...835....4B, Bhattacharyya:mdpi23}. Within this so-called 
propeller regime, the accelerated plasma may reach the escape speed and be ejected from the system. The expulsion of matter exerts a negative (braking) torque extracting angular momentum from the NS and causing it to spin down. Therefore, it is expected that accretion cannot spin up the NS beyond the 
point where $R_\mathrm{mag} = R_\mathrm{co}$. The fastest spin a NS can acquire from accretion is the 
so-called equilibrium spin period:
\begin{multline}
    \label{eq:p_eq}
    P_\mathrm{eq} \approx 0.26 \left(\frac{B}{10^8\,\mathrm{G}}\right)^{6/7} \left(\frac{R}{10\,\mathrm{km}}\right)^{18/7} \\ \times \left(\frac{M}{1.4\msun}\right)^{-5/7} \left(\frac{\mdot}{\mdot_\mathrm{Edd}}\right)^{-3/7}\,\mathrm{[ms]}
\end{multline}
\citep{10.1111/j.1365-2966.2005.09167.x}.

To calculate the resulting accretion torque as a function of time, we used the formulation suggested by \cite{Tauris:sc2012}:
\begin{equation}
    \label{eq:torque}
    N(t) = n(\omega) \left[\mdot(t) \sqrt{GM_\mathrm{ns}R_\mathrm{mag}(t)}\xi + \frac{\mu^2}{9R^3_\mathrm{mag}(t)}\right] - \frac{\dot{E}_\mathrm{dipole}(t)}{\Omega(t)},
\end{equation}
Equation \ref{eq:torque} describes the net torque governing the spin evolution of the NS. The first term represents the accretion torque. The second term corresponds to the magnetic torque arising from the interaction between the NS’s dipole field and the magnetosphere. The third term represents the spin-down torque caused by magnetic dipole radiation.
Here, $n(\omega)$ is a dimensionless quantity defined as $n(\omega) = \tanh \left( \frac{1 - \omega}{\delta \omega} \right)$, where $\omega$ is the fastness parameter given by $\omega = \sqrt{\left( \frac{R_\mathrm{mag}}{R_\mathrm{co}} \right)^3}$. The parameter $\xi$ is a factor of order unity and is fixed to $\xi=1$  in our calculations. The parameter $\mu$ denotes the magnetic dipole moment, $\dot{E}_\mathrm{dipole}(t)$ represents the time-dependent dipole energy loss, and $\Omega$ is the spin frequency.
The mass-transfer rate, $\mdot$, was taken directly from our \mesa binary stellar evolution models. The terms $G$, $M_\mathrm{ns}$, and $R_\mathrm{mag}$ represent the gravitational constant, NS mass, and magnetic radius, respectively.
The width of the transition zone near the magnetospheric boundary was assumed to be small enough ($\delta \omega = 0.02$) so that the quantity $n(\omega)$ corresponds to a step function, $n(\omega) = \pm 1$.
Finally, for a recycled pulsar, it is anticipated that the magnetic field decreases significantly during the early accretion phase, dropping by several orders of magnitude from its original strength \citep{1989Natur.342..656S}.
Because this reduction in the magnetic field is believed to occur over a timescale shorter than the duration of the mass-transfer phase \citep{Bhattacharya:1991pre, 2017ApJ...835....4B}, we considered the surface magnetic field strength to be  constant with a value of $B = 1.0 \times 10^8\,\mathrm{G}$. 

In practice, we used the temporal evolution of mass loss from the donor star $\dot{M}_{\rm d}(t)$ provided by \mesa, together with the mass-radius relations described above to calculate the evolution of the NS radius, spin and mass as a function of time. 
More specifically, the angular momentum $J_\mathrm{ns}$, was calculated as the discrete summation: 
\begin{equation}\label{eq:j_numerical_integration}
    J_\mathrm{ns}(t_n) = J_\mathrm{ns}(t_0) + \sum_{i=1}^n \Delta t_{i-1} N_{i-1},
\end{equation}
where the timestep $\Delta t$ was taken from our \mesa simulations, and the torque was evaluated using Eq.~\eqref{eq:torque}. Since the NS is considered old, we initialized the angular momentum as $J_\mathrm{ns}(0) = 0$.
 At each step, the radius, $R_\mathrm{ns}(M_\mathrm{ns}, J_\mathrm{ns})$, and spin frequency, $f_\mathrm{ns}(M_\mathrm{ns}, J_\mathrm{ns})$, were interpolated from our tabulated EoS data, using a linear radial basis function. 

 Although the temporal evolution of the NS mass was already provided by \mesa assuming that the NS accretes half of the mass lost by the donor (see Sect.~\ref{sec:input_params}), we reevaluated this parameter a posteriori using the torque model described in this section. More specifically, we assumed that mass accretion is limited by the Eddington mass accretion rate, and that the NS accumulates mass only when the fastness parameter is negative. 
 Due to this treatment, the NS mass evolves slightly differently than the accreting point mass as simulated with \mesa, with \textsc{m3} showing the largest difference, showing an increase in the average accretion efficiency of approximately 10\%.

\subsection{Accretion torque fluctuations} \label{sec:torque_fluctuations}
The torque equation, as expressed in Eq.~\eqref{eq:torque}, leads to fluctuations around the equilibrium frequency during the spin equilibrium phase, a phenomenon extensively reported in existing literature \citep[e.g.,][]{1997ApJS..113..367B, 2023arXiv231113303D}. For example, \cite{Tauris:sc2012} highlights how minor variations in the mass-transfer rate can impact the fastness parameter, resulting in rapid oscillations in the direction of the accretion torque. Similarly, \cite{1980ApJ...241L.155E} discuss the occurrence of alternating periods of spin-up and spin-down in pulsating X-ray sources, attributing these phases to variations in the luminosities of disk-fed NSs, which in turn affect the accretion torque. 
To assess the impact of these torque fluctuations on our findings, we conducted further analysis using additional \mesa models to explore the sensitivity of spin evolution to the timing resolution. We find that larger time intervals fail to accurately capture the Roche lobe decoupling phase, which interrupts the spin equilibrium. However, the evolution trajectories for spin, mass, and radius remained consistent across both larger and smaller time intervals. Consequently, even when adjusting the time resolution by an order of magnitude, the pre-transition mass is altered by less than 3\%. Therefore, we conclude that our findings are sufficiently robust against naturally occurring torque fluctuations.
However, studies of spin variations in numerous X-ray pulsar and accreting MSPs have raised a number of issues regarding the standard accretion model. Observational evidence of unexpected spin reversals and spin-down phases during outbursts in X-ray pulsars \citep[e.g.,][]{1997ApJS..113..367B, 2002ApJ...576L.137G} or accretion during the propeller phase challenge the conventional model used in this study and call for more detailed investigations on the accretion torque exerted on magnetized NSs.

\subsection{Orbital reconfiguration}\label{sec:orb_reconfig}
The evolution of the orbit and the NS spin were followed until the NS reached the phase transition threshold (see Fig.~\ref{fig:mvsr}). It was then assumed that the phase transition occurs instantaneously (i.e., on a timescale that is much shorter than the orbital period). To investigate the impact of the phase transition on the orbit, we used the prescriptions of \cite{1983apj:Hills} and \cite{Tauris:2017apj} in which  the post-transition semimajor axis is given as\begin{equation}
    \label{eq:major_axis_posttrans}
    \frac{a_\mathrm{f}}{a_\mathrm{i}} = \frac{1 - \Delta M/M}{1 - 2\Delta M/M - (w/v_\mathrm{rel})^2 - 2\cos\theta(w/v_\mathrm{rel})},
\end{equation}
where $\Delta M$ is the instantaneous mass defect  corresponding to the released gravitational binding energy during the phase transition, $M$ is the total mass of the pre-transition system, $v_\mathrm{rel}$ is the relative velocity between the two stars ($v_\mathrm{rel} = \sqrt{GM/a_i}$), $w$ is the magnitude of the kick velocity, and $\theta$ is the kick angle between the kick velocity vector, $\textbf{w}$, and the pre-transition orbital velocity vector.
The eccentricity of the post-transition binary system is given as
\begin{equation}
    \label{eq:eccentricity_posttrans}
    e = \sqrt{1 + \frac{2E_\mathrm{orb,f}L_\mathrm{orb,f}^2}{\mu_\mathrm{f} G^2 M_\mathrm{f,1}^2 M_\mathrm{f,2}^2}},
\end{equation}
where $L_\mathrm{orb,f} = a_\mathrm{i} \mu_\mathrm{f} \sqrt{(v_\mathrm{rel} + w\cos\theta)^2 + (w\sin\theta \sin\phi)^2}$ is the post-transition orbital angular momentum, with $\phi$ being the kick angle on the plane perpendicular to the pre-transition velocity vector, $\mu_\mathrm{f}$ is the post-transition reduced mass, and $E_\mathrm{orb,f} = -GM_\mathrm{f,1}M_\mathrm{f,2}/2a_\mathrm{f}$ is the post-transition orbital energy. 

To investigate the post-transition orbital configurations for our models, we considered a mass defect of $0.016\msun$ and secondary kicks with  magnitudes ($w$) up to $100$ km s$^{-1}$ (see Sect.~\ref{sec:mass_defect} for the justification). Furthermore, we assumed that the kick lacks a preferential orientation, leading us to model the kick angles $\theta$ and $\phi$ as uniformly distributed variables within a $4\pi$ solid angle.

\subsection{Mass defect and secondary kicks}\label{sec:mass_defect}
To provide a reliable estimate for the mass defect parameter in our simulations, we discuss here the estimates from the literature for the case of a ``catastrophic rearrangement'' due to a mass twin transition. According to \cite{Mishustin:2002xe}, the energy defect in the gravitational field of about $\Delta E \sim  7 \times 10^{51}$ erg can heat the star to an initial temperature of $T\simeq 40$ MeV. 
At this temperature, neutrinos are copiously produced and trapped inside the star because their mean free path is much smaller than the size of the star.
This leads to the establishment of a neutrino chemical potential $\mu_\nu$ in the direct and inverse $\beta$-decay processes, which establish the $\beta$-equilibrium including neutrinos. In a microscopic model for the quark de-confinement transition, the neutrino chemical potential influences on the critical density of the transition. 
Estimating the mass defect, one compares the gravitational masses of NSs in the initial and final states (before and after the transition) at a fixed baryon number. Without considering neutrino trapping, the mass defect for the EoS adopted here was calculated by \cite{2021AN....342..234A} to be in the range of 0.5 - 1.0\% for NS masses between $1.2$ and $1.6\msun$.
The mass defect due to the neutrino trapping--untrapping transition, modeled by a drop of the neutrino chemical potential from $150$ MeV to zero, has been found by \cite{Aguilera:2002dh} to be $\Delta M \approx 0.01\msun$. Due to a nonlinear dependence of $\Delta M$ on $\mu_\nu$ one can estimate the mass defect to be doubled for $\mu_\nu\approx 200$ MeV. 

In \cite{Aguilera:2002dh} a ``normal'' de-confinement transition without catastrophic rearrangement to a mass twin configuration has been considered. 
A scenario of neutrino untrapping in the cooling of a color superconducting quark star from a two-flavor, superconducting (2SC) phase to a three-flavor (Color-Flavor-Locked; CFL phase) mass-twin configuration has been examined in great detail for $\mu_\nu = 200$ MeV by \cite{Sandin:2007zr}. The untrapping transition led to a mass defect of $0.05\msun$ and the subsequent transition to the twin star with a CFL quark matter core produced a further mass defect of $0.03\msun$.
It has recently been shown that the mass defect associated with an accretion-induced thermal star quake can reach $3\%$ of a solar mass \citep{Carlomagno:2024vvr}.

\begin{figure}[t!]
    \centering
    \includegraphics[width=\columnwidth]{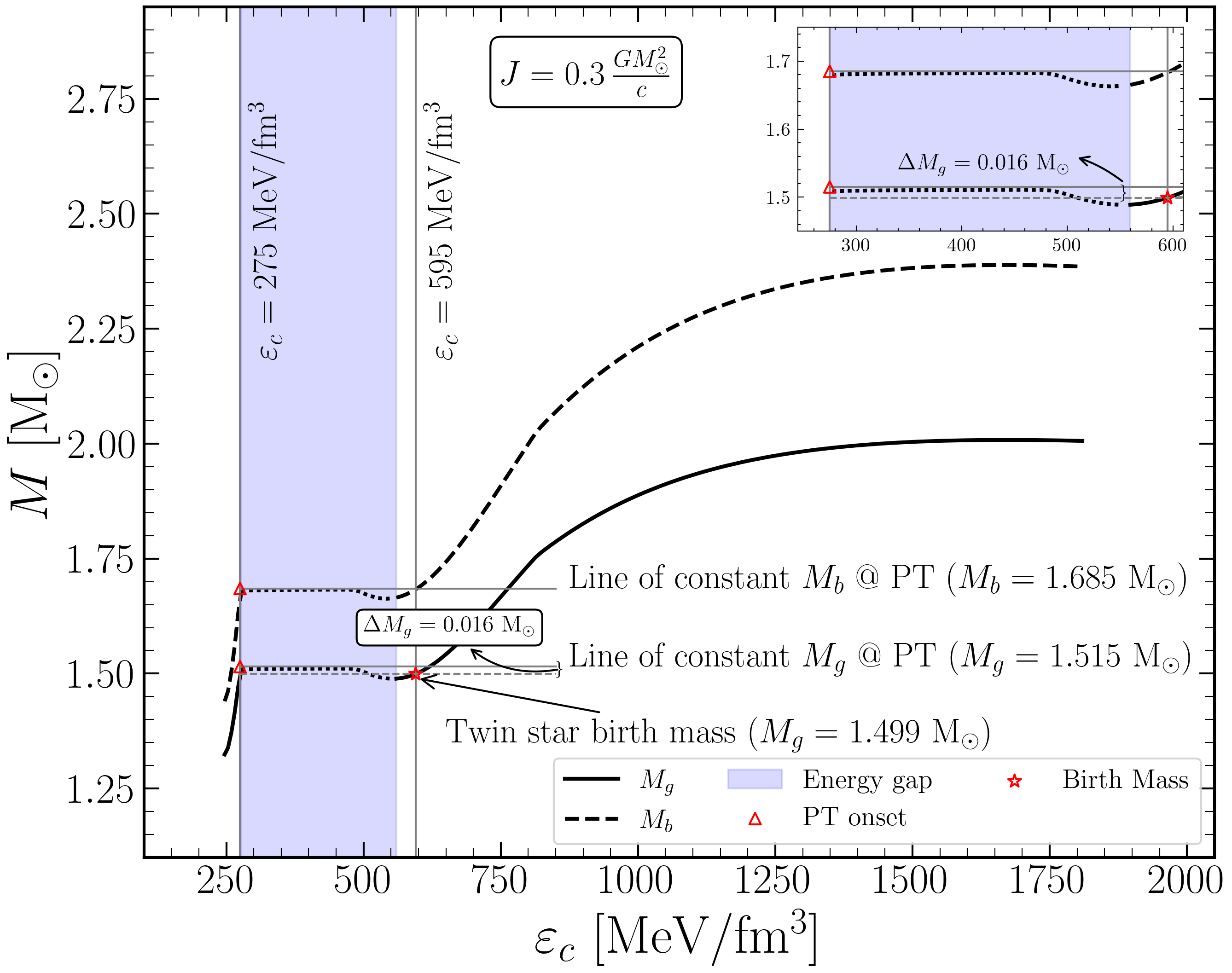}
    \caption{Baryonic mass ($M_b$) and gravitational mass ($M_g$) as functions of the central energy density ($\varepsilon_c$) for a fixed angular momentum ($J$). The solid black line represents the gravitational mass, while the dashed black line represents the baryonic mass. Vertical gray lines mark the central energy densities at the onset ($\varepsilon_c = 275\,\text{MeV/fm}^3$) and completion ($\varepsilon_c = 595\,\text{MeV/fm}^3$) of the phase transition (PT). Black dotted lines and the blue-shaded region denote unstable configurations.
    The inset zooms into the PT region, illustrating the mass defect ($\Delta M_g$) of $0.016\msun$ associated with the transition from hadronic to quark matter. The horizontal gray lines indicate constant values of $M_b$ and $M_g$ during the PT. The star marker denotes the birth mass of the twin star configuration at $M_g = 1.499\msun$.}
    \label{fig:mvse}
\end{figure}

Figure~\ref{fig:mvse} shows the baryonic mass (dot-dashed line) and gravitational mass (solid line) as functions of the central energy density for a fixed angular momentum, $J/J_0 = 0.3$. From this plot, the mass defect expected from a phase transition from hadronic to quark matter can be constructed by analyzing changes in these mass functions and identifying the critical points of the transition. By locating the central energy density at the onset of the transition and noting that the baryonic mass remains constant, one can determine the energy jump and identify the central energy density marking the completion of the transition. Tracing the corresponding gravitational mass at these points allowed us to compare the gravitational mass before and after the transition. The difference between these values represents the mass defect, which quantifies the mass converted into energy during the phase transition.

\begin{figure*}[thb]
    \centering
    \includegraphics[width=\textwidth,angle=0]{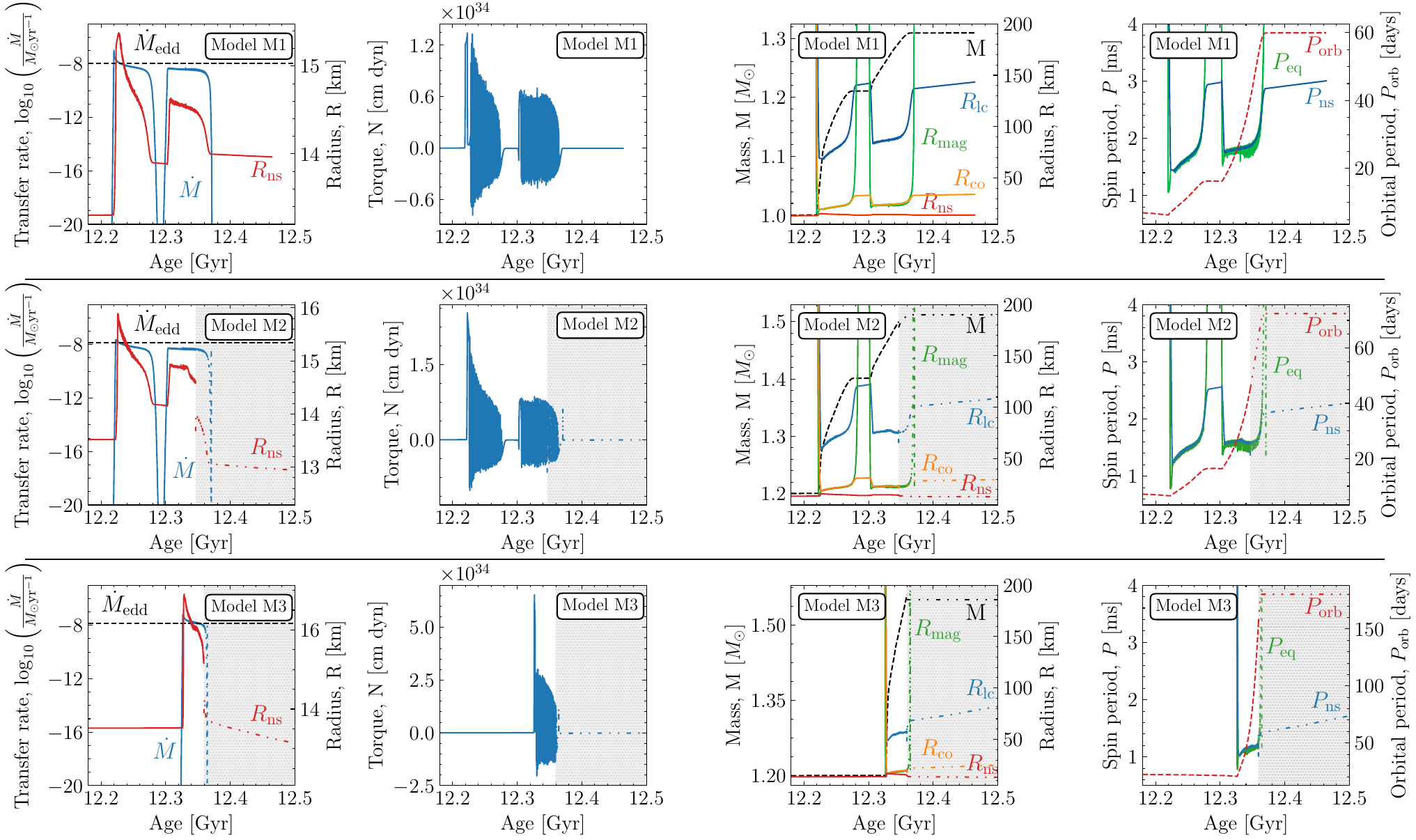}
    \caption{Time evolution of \textsc{m1} (top panel), \textsc{m2} (middle panel), and \textsc{m3} (bottom panel). The first column shows the evolution or mass-transfer rate (left y-axis) and NS radius (right y-axis). The second column shows the accretion torque. The third column shows the NS mass (left y-axis,  dashed black line) and various radii (right y-axis: magnetic, corotation, light-cylinder, and NS radius). The fourth column shows the spin period (left y-axis) and orbital period (right y-axis). Gray regions indicate post-transition evolution, with non-solid tracks for \textsc{m2} and \textsc{m3}.
    For a detailed description of the three spin evolution scenarios, see the main text.}
    \label{fig:lmxb_grid}
\end{figure*}   

Our present EoS in the form of a multi-polytrope model is too simple to enable a more precise quantitative estimate of the mass defect  for a realistic dense-matter model with neutrino untrapping and (three-flavor) color superconductivity. This task is left to separate work.
In any case, the estimates provided in this subsection justify a variation of the mass defect parameter in a range $0.1\%$ up to $5\%$ assumed in \cite[see also \citealt{Jiang:raa2021}]{Jiang:apj15}.

Besides orbital rearrangement due to a mass defect, we also considered the possibility of a secondary kick during the transition \citep{Hobbs:2005yx,Bombaci:2004nu}. 
For the range of mass defects considered here, anisotropic emission could give rise to secondary kicks with magnitudes of up to $w \simeq 20,000$\,km\,s$^{-1}$.
One possible mechanism that can give rise to such anisotropies is the channeling of the neutrino emission along a preferential axis formed by magnetic vortex lines together with parity violation in the strong internal magnetic field of the NS  \citep[see][for recent examples with typical B-field strengths of $B\sim 10^{12}$ Gauss]{Berdermann:2006rk,Kaminski:2014jda,Fukushima:2024cpg}.

The electromagnetic rocket effect, which was introduced in \cite{1975ApJ...201..447H} and revisited by \cite{Agalianou:2023lvv}, is based on a displacement of a similarly strong dipolar magnetic field from the center of the NS and its sufficiently fast rotation.
As it was discussed in \cite{Agalianou:2023lvv}, observational evidence provided by the Neutron Star Interior Composition Explorer (NICER) X-ray observatory, particularly from investigating the location of hot spots on the surfaces of MSPs \citep{Miller:2019cac, Riley:2019yda}, indicates that the magnetic field diverges from a centered dipole configuration.
 Based on the results for PSR J0030+0451, \cite{Kalapotharakos:2020rmz} estimated the magnetic field structure and found an off-centered magnetic field consisting of dipole and quadruple components. 

Since both the neutrino and electromagnetic rocket mechanisms require strong magnetic fields of at least the young pulsar field strength $B\sim 10^{12}$ Gauss, such effects seem at first glance unlikely for ``old'' MSPs with low magnetic fields of
$B\sim 10^{8}$ Gauss.
However, in systems with mass accretion from a companion star like in the case of LMXBs, the argument of \cite{1989ApJ...346..847C} can be applied, that the strong magnetic field of the interior gets buried in the crust of the pulsar \citep[e.g.,][and references therein]{Bhattacharya:1991pre, 2001ApJ...557..958C, 10.1111/j.1365-2966.2004.07397.x}.
Thus, a small surface magnetic field is compatible with a super-strong magnetic field in the interior. 
Due to such a magnetic field profile, kick velocities of a few dozen km s$^{-1}$ can be produced in MSPs, as it has been demonstrated by \cite{Agalianou:2023lvv} by applying the electromagnetic rocket effect to the case of PSR J0030+0451.

A gravitational recoil effect arising from the anisotropic redistribution of matter following the phase transition, could also produce a secondary kick. This effect is independent of the magnetic field strength, making it a viable secondary kick mechanism for pulsars across a wide range of magnetic field intensities, including old MSPs with weak magnetic fields.
Considering the above, we included a possible secondary kick mechanism in our study by parametrically varying the kick velocity parameter in the range $w=0, \dots, 100$ km s$^{-1}$.

\section{Results} \label{sec:results}
\subsection{Mass and spin accretion} \label{sec:pretrans_evolution}
Figure~\ref{fig:lmxb_grid} illustrates the evolution of our binary models from approximately the onset of the LMXB phase. The top and middle panels showcase that, due to their comparable initial 
setups,  \textsc{m1} and \textsc{m2} have similar characteristics. 
Initially, the donor star in each model spends most of its life in the main sequence. After 
approximately $12.2\,\text{Gyr}$, it reaches the tip of the red giant branch and it fills its Roche 
lobe, initiating a mass flow toward the NS with a mass-transfer rate of $\mdot \simeq 
10^{-9}\mdotsun$. As detailed in Sect.~\ref{sec:methods}, the interaction of the accreted 
matter with the magnetic field of the NS generates an accretion torque \citep{Tauris:sc2012}. We find that the exerted torque significantly decreases the NS spin to nearly 
$1.1\,\rm{ms}$ shortly after the RLOF is initiated.
Over the next circa $56\,\rm{Myr}$ the mass-transfer rate gradually decreases, allowing the spin of the NS to approach a stable equilibrium period, exhibiting only minor fluctuations due to rapid 
torque reversals. 
At this point, the binary orbit widens to a period of $P_\mathrm{orb} \simeq 16.5\,\mathrm{d}$ with the 
donor star separating from its Roche lobe after shedding nearly half of its total mass. In contrast, 
the NS has gained approximately $0.2\msun$.
As the mass-transfer rate decreases, so does the ram pressure, allowing the magnetospheric boundary to 
extend outward. Initially, during the early stages of Roche lobe decoupling, the NS spin 
decreases while remaining in equilibrium. However, the reduction in the mass-transfer rate over time 
disrupts the equilibrium, allowing the NS to enter a relatively brief phase, lasting 
approximately $28\,\rm{Myr}$, where its rotation rate further decreases. Eventually, the pulsar's spin 
relaxes to an average value of $\langle P_{\rm ns} \rangle = 2.8\,\rm{ms}$. Following this, the donor 
star refills its Roche lobe, initiating a second episode of mass transfer that lasts for approximately 
another $\simeq 63\,\text{Myr}$. In this stage, the NS mass increases, accompanied by a 
steady contraction in radius, ultimately reaching approximate values of $1.3\msun$ and 
$14\,\text{km,}$ respectively, for \textsc{m1}. Simultaneously, the donor star transitions into a 
low-mass helium WD with $M_d \simeq 0.29\msun$.
In all cases, during the late stages of the evolution, the magnetospheric radius surpasses the light-cylinder radius. When this happens, the evolution of the NS is only driven by the magnetic dipole radiation torque (i.e., the third term of Eq.~\eqref{eq:torque})

In contrast, toward the end of this second mass-transfer phase, \textsc{m2} surpasses the maximum mass that can be supported by pure hadronic matter. This event is visually represented in Fig.~\ref{fig:mvsr} when the green curve intersects the phase transition boundary. At this point, the NS mass is approximately $1.5\msun$, and it has a radius of about $14.6\,\mathrm{km}$, while the orbital period of the binary system is $45.2\,\mathrm{d}$. To model the NS's core transitioning from hadronic to quark matter due to accretion, we assumed a minor loss in gravitational mass of $\Delta M = 0.016\msun$. This assumption requires the core's instantaneous collapse, with both angular momentum ($J$) and baryonic mass ($M_b$) conserved, yielding a denser object, now with a reduced radius of approximately $13.7\,\mathrm{km}$. This transition is indicated by the black arrow in Fig.~\ref{fig:mvsr}, where its position on the third-family branch for compact objects is marked with a green star. For \textsc{m2}, the twin star is 6.2\% more compact than its purely hadronic counterpart.

\begin{figure*}[th!]
    \centering
    \includegraphics[width=0.45\textwidth]{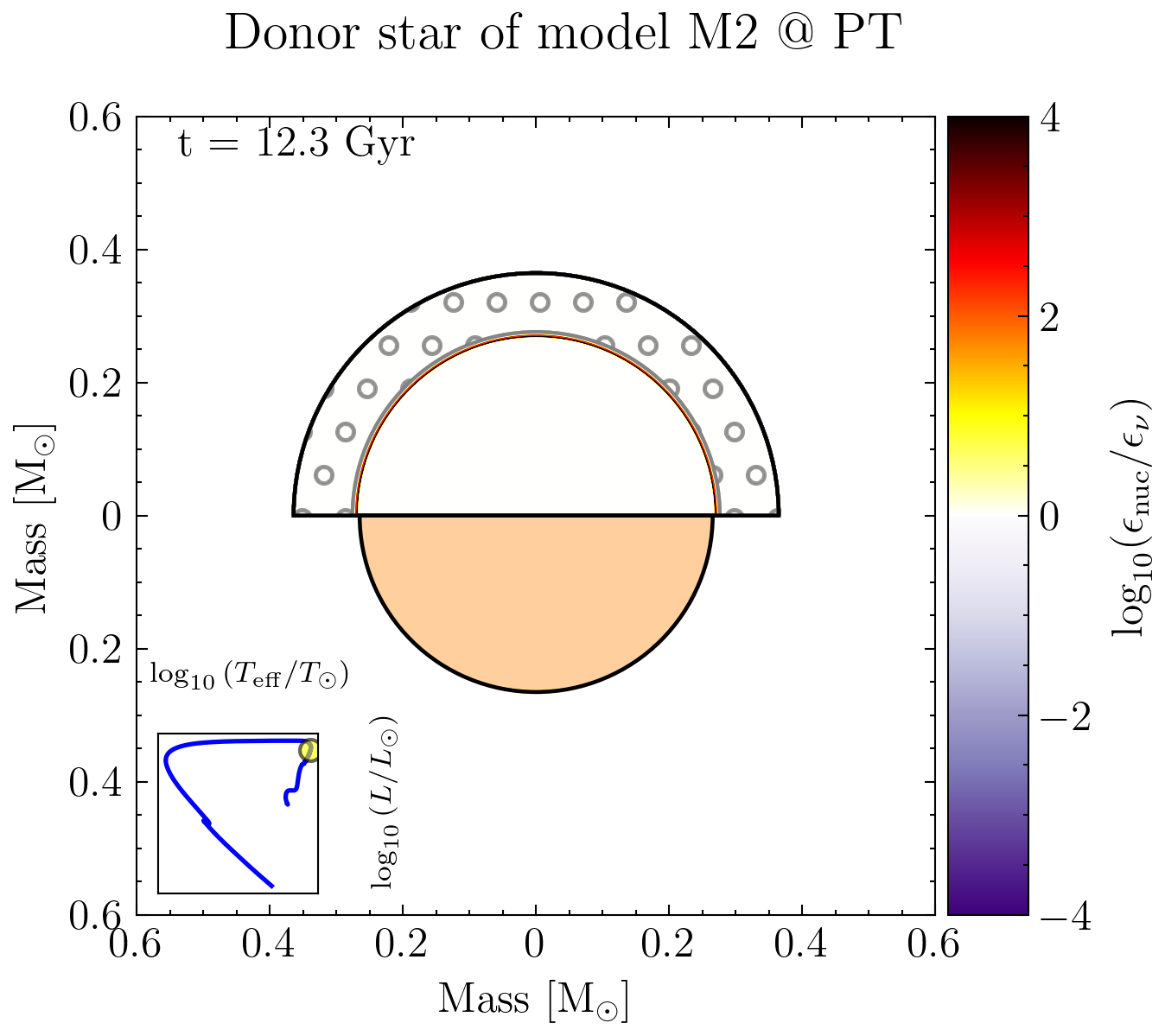}
    \hspace{1cm}
    \includegraphics[width=0.45\textwidth]{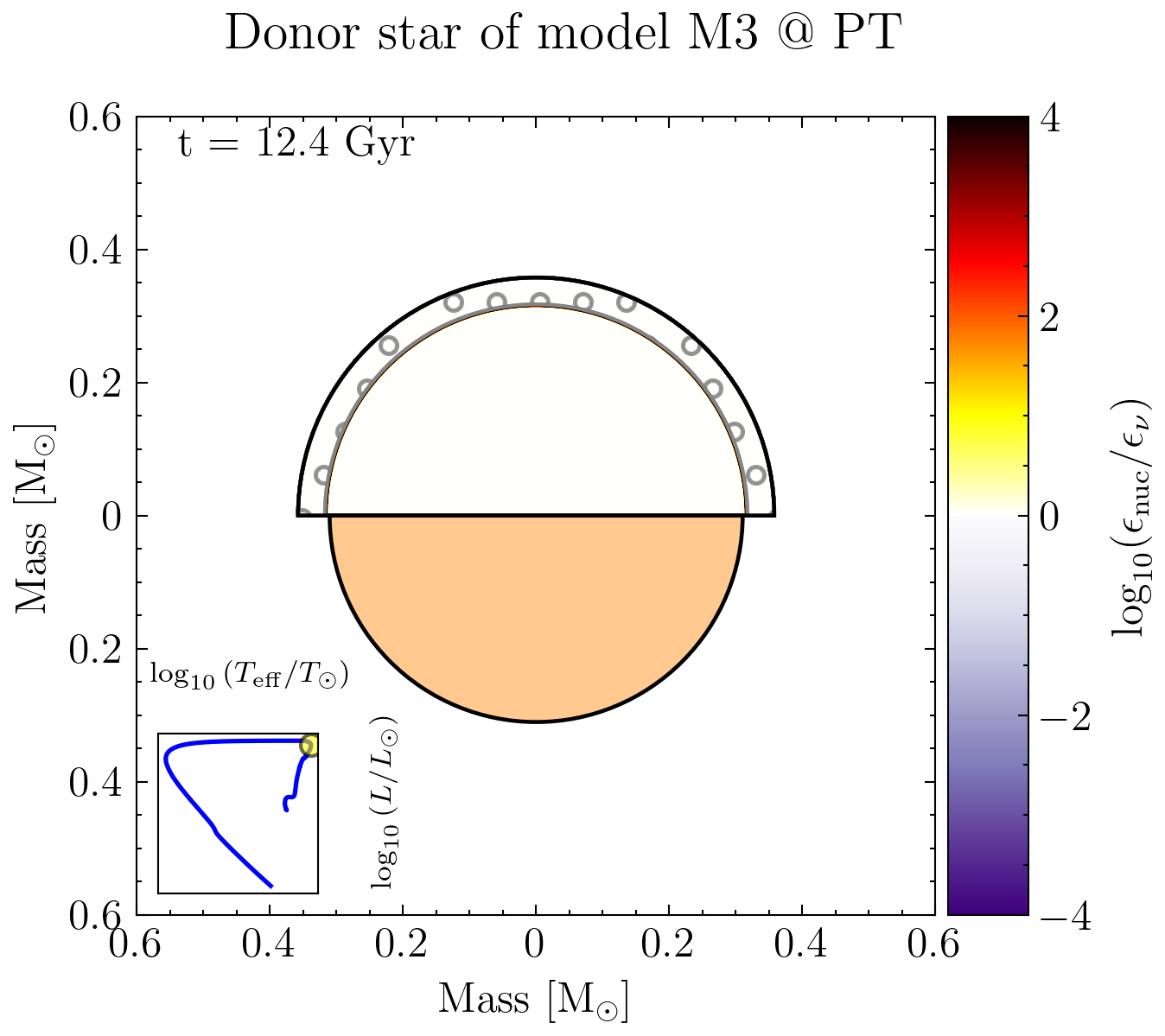}
    \caption{Structure of the donor star. The top half of each figure shows the energy generation and interior mixing of the donor star at the time of the phase transition for \textsc{m2} (left) and \textsc{m3} (right). The bottom half shows the helium core mass and the color of the star based on its surface temperature. The yellow circle in the inline plot shows their position on a standard Hertzsprung-Russell diagram; the x-axis is the logarithm of the effective temperature, and the y-axis is the logarithm of surface luminosity. Dotted areas depict regions where convection is the dominant mixing mechanism.}
    \label{fig:donor_plots}
\end{figure*}

The structure of the donor star at the time of the phase transition is shown on the left panel of Fig.~\ref{fig:donor_plots}. The bottom half shows the helium core mass, which is approximately equal to $0.27\msun$, and the perceived color of the star (as seen by an observer) based on its surface effective temperature. The inline plot illustrates the evolution of the donor star on the Hertzsprung-Russell diagram with the yellow circle marking its current position. The top half depicts the energy generation that takes place on a thin layer above the helium core with the color bar indicating the energy generation rate divided by the losses due to neutrino emission.
The dominant mixing processes can also be discerned; at this point, the donor star has a fully convective envelope with a mass of approximately $0.1\msun$ and a  radius of $\sim 17\rsun$. 

Binary systems harboring rapidly rotating NSs in wider orbits, such as in \textsc{m3}, 
can experience a delayed accretion-induced phase transition as a result of the spin-down phase
following the cessation of mass transfer (see the blue curve in Fig.~\ref{fig:mvsr}). Owing to the 
longer initial orbital period, \textsc{m3} experiences a single mass-transfer episode 
due to a reduced gravitational influence from the NS, allowing the donor star to maintain a more stable evolutionary path. As a consequence, the donor star does not expand significantly until it reaches a later stage in its evolution, where it fills its Roche lobe only once.
The mass transfer commences approximately $104\,\mathrm{Myr}$ later compared to the first RLOF for \textsc{m1} and \textsc{m2} (bottom panel in Fig.~\ref{fig:lmxb_grid}). The duration of the
mass transfer is close to $33\,\mathrm{Myr}$ and, at the end of it, the donor and the NS possess a mass of approximately $0.33\msun$ and $1.56\msun$, respectively. Subsequent to Roche lobe detachment, the NS spin rate decreases for approximately $117\,\mathrm{kyr}$ at which point becomes unstable and undergoes a phase transition. Similarly to \textsc{m2}, its trajectory and final position on the third-family branch are marked with a black arrow and a blue star, respectively. The resulting twin star is $\sim 8.4\%$ more compact than the initial NS, settling on a radius of approximately $13.9$ km. 

The structure of the donor star at the time of the phase transition can be seen in Fig.~\ref{fig:donor_plots}. Although similar to the structure of the donor star in \textsc{m2}, the envelope mass is lower ($\sim 0.04\msun$) and its total radius almost twice that of \textsc{m2} ($\sim 34\rsun$) as it enters the asymptotic giant branch phase.

\begin{figure*}[t]
    \centering
    \includegraphics[width=0.45\textwidth]{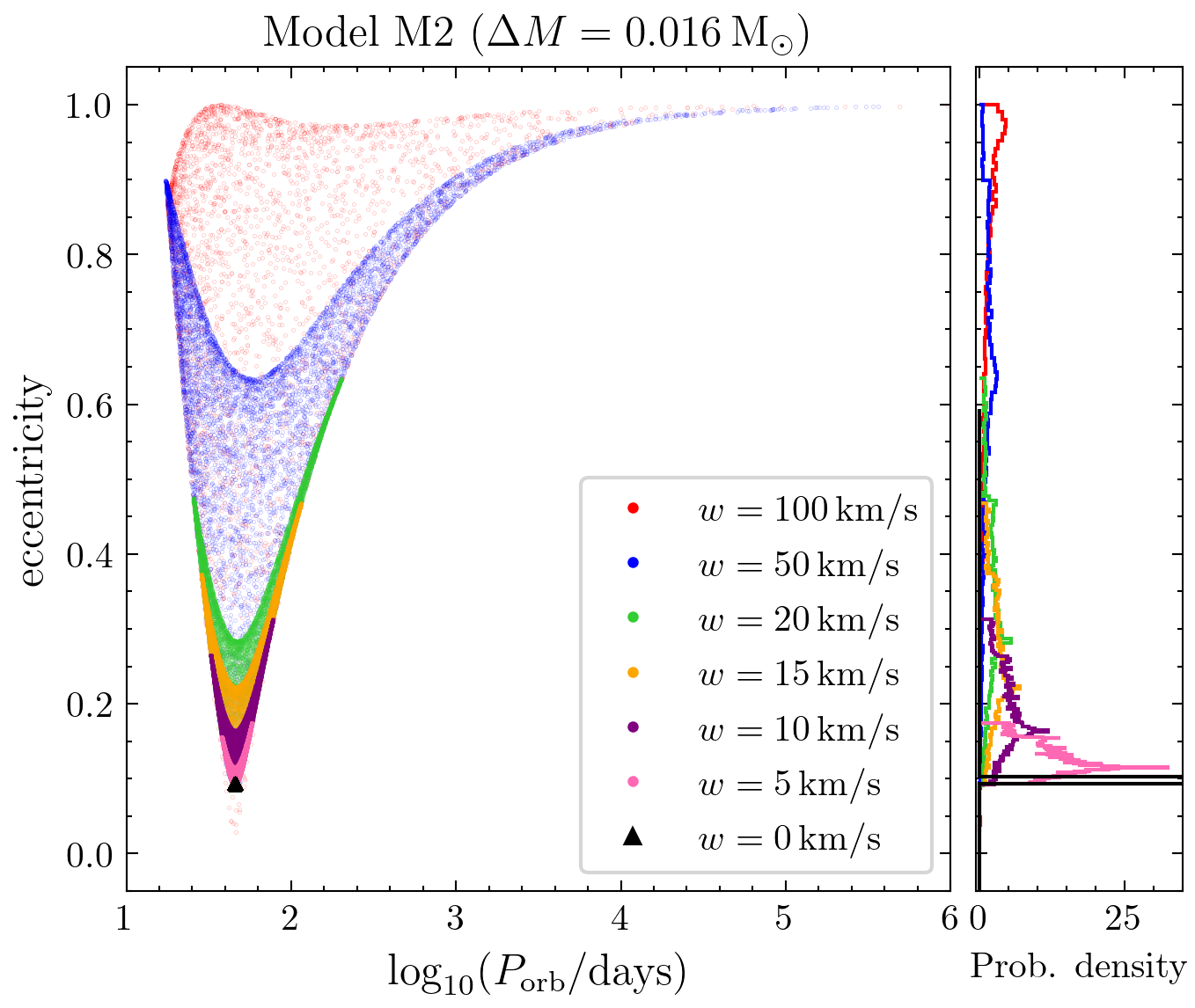}
    \hspace{1cm}
    \includegraphics[width=0.45\textwidth]{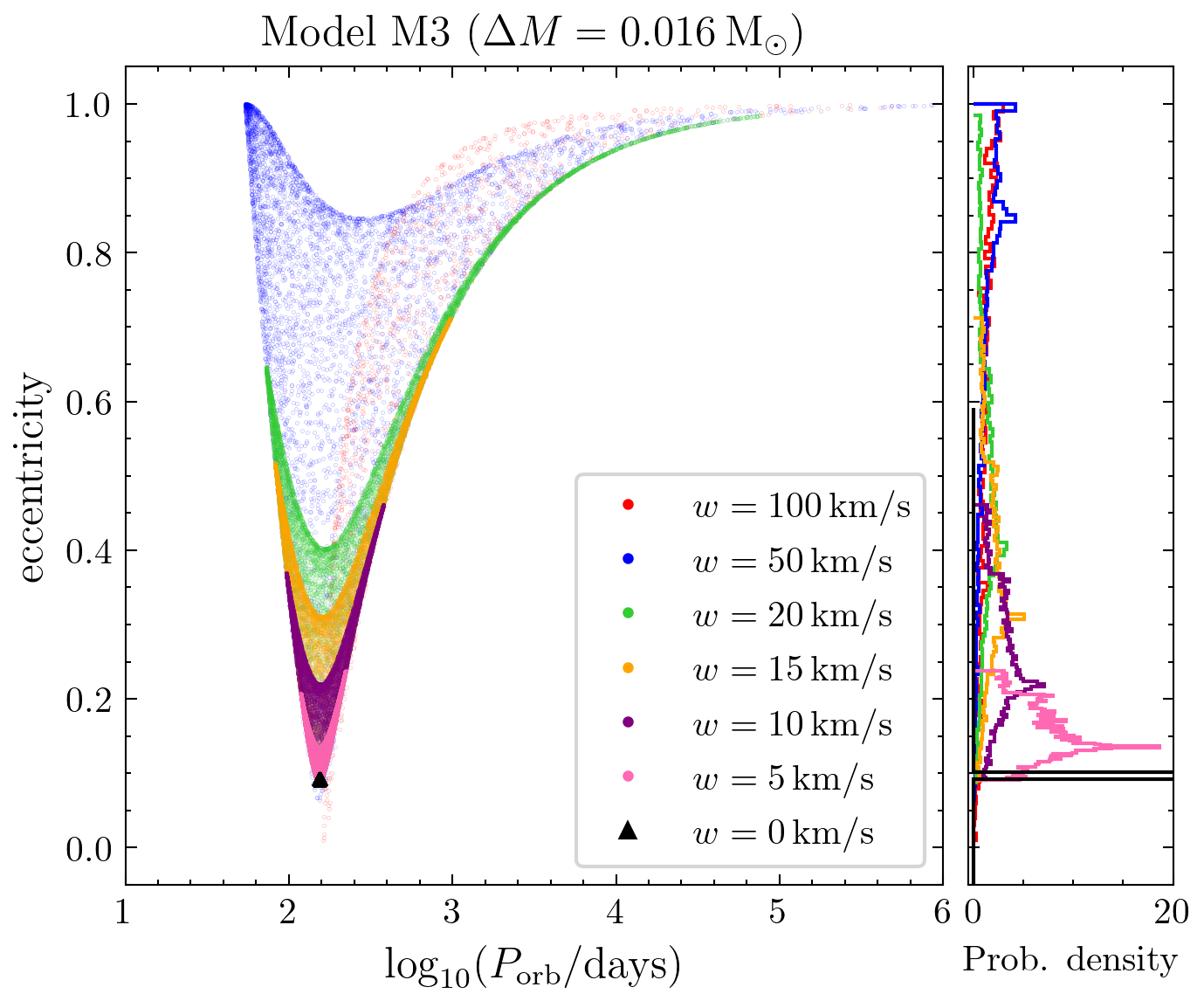}
    \caption{Monte Carlo simulation for the post-transition distribution of orbital parameters from\textsc{m2} (left) and \textsc{m3} (right). A transition-triggered mass loss of $\Delta M = 0.016\msun$ and a fixed magnitude kick velocity were assumed. The kick orientation was randomly sampled from a uniform distribution.}
    \label{fig:mc_sim}
\end{figure*}

\subsection{Orbital reconfiguration and post-transition evolution} \label{sec:posttrans_evolution}
Figure~\ref{fig:mc_sim} illustrates the post-transition binary system configurations for our progenitor models, \textsc{m2} and \textsc{m3,} for kick velocities with fixed magnitudes and random orientations imparted on the compact star. 
For both \textsc{m2} and \textsc{m3},  even in the absence of a kick ($w=0\,\mathrm{km\,s^{-1}}$; depicted by the black markers), the binary systems exhibit an increase in their eccentricity that is directly proportional to the mass defect during the mass transition, $e = \Delta M / M \simeq 7\times10^{-2}$, for $\Delta M = 0.016\msun$. 
For $\textsc{m3}$, these orbital changes are preserved, as the system does not experience additional mass transfer after the phase transition, and thus the tidal forces are negligible \citep{Freire:mnras14, antoniadis:apjl14, Jiang:apj15}. 
When factoring in a secondary kick mechanism, as the kick velocity increases, there is a progressively larger spread in the distribution of both eccentricity and binary separation, leading to a more diverse range of possible post-kick orbits. 
For a kick of certain magnitude, wider systems such as \textsc{m3} are more likely to produce more eccentric systems or eject the compact star, resulting in the formation of an eMSP or iMSP, respectively.  

Evolutionary calculations following the phase transition, provided by \mesa, in \textsc{m2} may be unreliable as they do not take into account the effect of the orbital reconfiguration on the mass transfer. In contrast, the simulation for \textsc{m3}, where the phase transition occurs during Roche lobe decoupling, remains generally reliable. However, it is important to note that the final orbital period in \textsc{m3} may still be inaccurate due to the limitations in capturing the orbital reconfiguration.
In addition, the radius of the hybrid star decreases as a result of the phase transition. Due to the conservation of angular momentum, this reduction leads to a 15\% increase in the spin frequency of the compact star. 
Moreover, because of the magnetic flux conservation during the phase transition, the surface magnetic field of the hybrid star also increases. The new magnetic field strength can be calculated as
\begin{equation}
    B_\mathrm{hybrid} = B_\mathrm{ns} \left(\frac{R_\mathrm{ns}}{R_\mathrm{hybrid}}\right)^2 \simeq 1.14 \times 10^8\,\mathrm{G},
\end{equation}
where $B_\mathrm{ns}$ and $R_\mathrm{ns}$ are the magnetic field and the radius of the NS before the transition, respectively, and $R_\mathrm{hybrid}$ is the radius of the hybrid star after the transition.

For \textsc{m2} where the phase transition takes place during the LMXB phase, the magnetospheric radius outgrows the corotation radius (i.e., the fastness parameter becomes greater than one) forcing the hybrid star to enter a propeller phase and inhibit further accretion. Hence, it is conceivable that the NS mass stops increasing despite the ongoing mass transfer. Moreover, if the magnetospheric radius outgrows the light cylinder radius, the magnetic field is strong enough to prevent accretion altogether. This may lead to the formation of a binary system that consists of a rotation-powered pulsar and an evolved ``spider''-like companion \cite[see][and references therein]{Strader:2019,Papitto:2022}.  

Although we did not perform detailed calculations with \mesa for the post-transition evolution of \textsc{m2} and \textsc{m3}, we used the same accretion model to compute the mass, radius, and spin evolution of the hybrid stars (dash double dotted lines in Fig.~\ref{fig:lmxb_grid}). We find that \textsc{m2} continues to go through sporadic episodes of mass accretion reaching a final mass of approximately $1.51\msun$. 

In contrast, \textsc{m3}, which had already decoupled from its Roche lobe prior to the strong phase transition, retained its post-transition mass of approximately $1.55\msun$. Following the cessation of  mass transfer, both models are further compactified as a result of the spin-down and angular momentum loss (see also Fig.~\ref{fig:mvsr}). By the end of the evolution, the hybrid star of \textsc{m2} has a radius of approximately $12.4$ km compared to its pure hadronic counterpart with a radius of $14.6$ km at the time of transition. Similarly, in \textsc{m3}, the hybrid star has a radius of $12.2$ km whereas the pure hadronic star has a radius of $15.2$ km.

\section{Discussion} \label{sec:summary}

\subsection{Summary}
In this work we investigated the potential role of rapid first-order phase transitions from ordinary baryonic to quark matter in the evolution of LMXB and post-LMXB systems. 
We performed detailed binary evolution modeling using \mesa and Monte Carlo simulations to explore the influence of mass transfer, accretion- and spin-down-induced phase transitions, and secondary kick mechanisms on the formation and evolution of these systems. Our results demonstrate that:
\begin{enumerate}
    \item Twin stars can form as a result of mass transfer in LMXBs 
    \item In systems with sufficiently long orbital periods, strong phase transitions can occur after the LMXB phase, when the NS is a rotation-powered MSP. In this case, the eccentricity of the resulting system can be several orders of magnitude larger than what is expected for recycled MSPs \citep{Phinney:1992}.
    \item If the transition is accompanied by a secondary kick (see Sect.~\ref{sec:orb_reconfig} for the justification), then a considerable number of systems can be disrupted, even if the kick magnitude is moderate. Therefore, this mechanism may be a viable formation path for iMSPs.    \item Secondary kicks also provide a mechanism for producing binary MSPs with very long orbital periods ($>1000$\,d) from systems that are initially more compact. 
\end{enumerate}

\subsection{Formation of isolated and eccentric millisecond pulsars}\label{sec:msps}

As discussed in Sect.~\ref{sec:posttrans_evolution},  the minimum post-transition eccentricity is directly related to the mass defect, $\Delta M$. The EoS model adopted in this work predicts a mass defect of $\Delta M = 0.016\msun$, which in turn produces eccentricities  of $e\leq\Delta M / M \simeq 7 \times 10^{-2}$. 

For \textsc{m2} this residual eccentricity would rapidly dissipate as a result of tidal forces acting on the envelope of the donor star \citep{1977A&A....57..383Z}.  
In contrast, for \textsc{m3} the strong phase transition occurs during the spin-down phase (i.e., after the LMXB phase), when the donor star is already evolving toward the WD cooling track. For this reason, the resulting orbital changes would persist \citep{Freire:mnras14,antoniadis:apjl14}, providing a potential mechanism for explaining eMSPs.  Although this particular model has an orbital period that is too long compared to those of known eMSPs with WD companions, it shows that strong phase transitions provide a possible formation mechanism for eMSPs.

Because high-precision pulsar timing can constrain extremely small eccentricities ($\leq 10^{-8}$), even models with no neutrino trapping and with small mass defects that would produce smaller deviations from the \cite{Phinney:1992} relation could potentially be observed. Our work suggests that the  MSP eccentricity distribution, especially in binary MSPs with long orbital periods,  can be used to constrain  not only stellar physics \citep{Phinney:1992, antoniadis:apjl14}, but also  nuclear matter models.

\begin{figure}[t]
    \centering
    \includegraphics[width=0.45\textwidth]{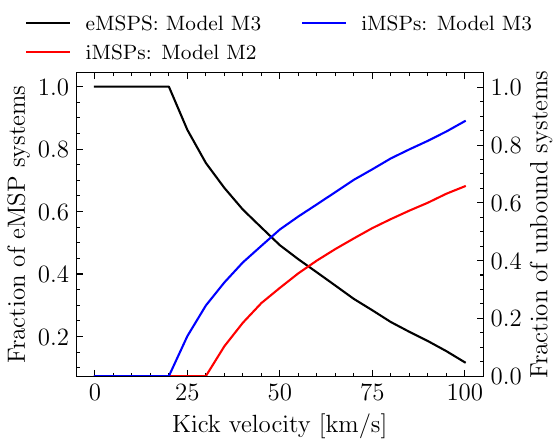}
    \caption{Expected fraction of eMSPs (black curve, left y-axis) and iMSPs (red and blue curves, right y-axis) as a function of kick velocity, assuming a mass defect of $\Delta M = 0.016\msun$. For iMSPs, we consider the systems with eccentricities greater than unity. \textsc{m2} cannot create eccentric systems due to the rapid re-circularization of the orbit.}
    \label{fig:emsps_vs_imsps}
\end{figure}

When considering secondary kick mechanisms, higher kick velocities produce a broader distribution of eccentricities and orbital periods, suggesting a diverse range of possible post-transition configurations. Figures~\ref{fig:mc_sim} and \ref{fig:emsps_vs_imsps} illustrate that both \textsc{m2} and \textsc{m3}  have a high probability of producing iMSPs or binary MSPs with extremely long orbital periods. More specifically, the red and blue curves in Fig.~\ref{fig:emsps_vs_imsps} (representing iMSPs from  \textsc{m2} and \textsc{m3}, respectively) illustrate the expected fraction of binary systems that, due to the imparted kick velocities, achieve eccentricities greater than unity — indicating the complete disruption of the binary and the formation of an iMSP. As the kick velocity increases, a larger fraction of systems become unbound. In contrast, the black curve on the left y-axis, representing the fraction of eMSPs in \textsc{m3}, shows an inverse relationship: as the kick velocity increases, the fraction of eMSPs declines. This suggests that higher kick velocities are more likely to disrupt the binary systems entirely, thereby preventing the formation of eMSPs.

In our study we considered three different mechanisms that could introduce such a kick: the neutrino rocket effect, the electromagnetic rocket effect, and the GW recoil scenario. In the last case, even if a small percentage ($\sim 0.5\%$) of the converted energy is emitted in the form of GWs, it can account for the desired kick velocities. However, a fully relativistic approach to modeling the GW emission is necessary to accurately quantify the results of such a recoil.
Nevertheless, the possibility that these three mechanisms could coexist, or sequentially influence the binary during various evolutionary stages, is intriguing since each contributes differently to the system dynamics.  
Future pulsar surveys that are more sensitive to these systems  will therefore play an important role in constraining the potential birth signatures of twin stars \citep{watts:2015, antoniadis:2021}.

\begin{acknowledgements}
This work was supported by the Stavros Niarchos Foundation (SNF) and the Hellenic Foundation for Research and Innovation (H.F.R.I.) 
under the 2nd Call of ``Science and Society'' Action Always strive for excellence -- ``Theodoros Papazoglou’' (Project Number: 01431).
S.Ch and J.A. also acknowledge support from the European Commission under project ARGOS-CDS (Grant Agreement number: 101094354).
D.B. acknowledges support from the Polish National Science Center under grant No. 2019/33/B/ST9/03059 (before 10/2023) and under grant No. 2021/43/P/ST2/03319 (after 04/2024). D.E.A-C. is supported by the program Excellence Initiative–Research University of the University of Wroclaw of the Ministry of Education and Science. This work made extensive use of the following software:
\texttt{Astropy} \citep{2013A&A...558A..33A,2018AJ....156..123A}, \texttt{NumPy} \citep{harris2020array:numpy}, \texttt{SciPy} \citep{2020SciPy-NMeth}, \texttt{Pandas} \citep{reback2020pandas}, \texttt{Matplotlib} \citep{Hunter:2007:matplotlib}, \texttt{SciencePlots} \citep{SciencePlots}, \texttt{MesaReader} \citep{pymesareader}, \texttt{Jupyter} \citep{Kluyver2016jupyter}. Figure~\ref{fig:donor_plots} was created with \texttt{TULIPS}\footnote{\url{https://astro-tulips.readthedocs.io/en/latest/}} \citep{2022A&C....3800516L}.
\end{acknowledgements}


\bibliographystyle{aa}
\bibliography{bibfile}

\begin{thebibliography}{111}
\expandafter\ifx\csname natexlab\endcsname\relax\def\natexlab#1{#1}\fi

\bibitem[{Abbott {et~al.}(2017)Abbott, Abbott, Abbott, Acernese, Ackley, Adams,
  Adams, Addesso, Adhikari, Adya, Affeldt, Afrough, Agarwal, Agathos, Agatsuma,
  Aggarwal, Aguiar, Aiello, Ain, Ajith, Allen, Allen, Allocca, Altin, Amato,
  Ananyeva, Anderson, Anderson, Angelova, Antier, Appert, Arai, Araya, Areeda,
  Arnaud, Arun, Ascenzi, Ashton, Ast, Aston, Astone, Atallah, Aufmuth, Aulbert,
  AultONeal, Austin, Avila-Alvarez, Babak, Bacon, Bader, Bae, Bailes, Baker,
  Baldaccini, Ballardin, Ballmer, Banagiri, Barayoga, Barclay, Barish, Barker,
  Barkett, Barone, Barr, Barsotti, Barsuglia, Barta, Barthelmy, Bartlett,
  Bartos, Bassiri, Basti, Batch, Bawaj, Bayley, Bazzan, B\'ecsy, Beer, Bejger,
  Belahcene, Bell, Berger, Bergmann, Bernuzzi, Bero, Berry, Bersanetti,
  Bertolini, Betzwieser, Bhagwat, Bhandare, Bilenko, Billingsley, Billman,
  Birch, Birney, Birnholtz, Biscans, Biscoveanu, Bisht, Bitossi, Biwer,
  Bizouard, Blackburn, Blackman, Blair, Blair, Blair, Bloemen, Bock, Bode,
  Boer, Bogaert, Bohe, Bondu, Bonilla, Bonnand, Boom, Bork, Boschi, Bose,
  Bossie, Bouffanais, Bozzi, Bradaschia, Brady, Branchesi, Brau, Briant,
  Brillet, Brinkmann, Brisson, Brockill, Broida, Brooks, Brown, Brown, Brunett,
  Buchanan, Buikema, Bulik, Bulten, Buonanno, Buskulic, Buy, Byer, Cabero,
  Cadonati, Cagnoli, Cahillane, Calder\'on~Bustillo, Callister, Calloni, Camp,
  Canepa, Canizares, Cannon, Cao, Cao, Capano, Capocasa, Carbognani, Caride,
  Carney, Carullo, Casanueva~Diaz, Casentini, Caudill, Cavagli\`a, Cavalier,
  Cavalieri, Cella, Cepeda, Cerd\'a-Dur\'an, Cerretani, Cesarini, Chamberlin,
  Chan, Chao, Charlton, Chase, Chassande-Mottin, Chatterjee, Chatziioannou,
  Cheeseboro, Chen, Chen, Chen, Cheng, Chia, Chincarini, Chiummo, Chmiel, Cho,
  Cho, Chow, Christensen, Chu, Chua, Chua, Chung, Chung, Ciani, Ciolfi,
  Cirelli, Cirone, Clara, Clark, Clearwater, Cleva, Cocchieri, Coccia, Cohadon,
  Cohen, Colla, Collette, Cominsky, Constancio, Conti, Cooper, Corban, Corbitt,
  Cordero-Carri\'on, Corley, Cornish, Corsi, Cortese, Costa, Coughlin,
  Coughlin, Coulon, Countryman, Couvares, Covas, Cowan, Coward, Cowart, Coyne,
  Coyne, Creighton, Creighton, Cripe, Crowder, Cullen, Cumming, Cunningham,
  Cuoco, Dal~Canton, D\'alya, Danilishin, D'Antonio, Danzmann, Dasgupta,
  Da~Silva~Costa, Dattilo, Dave, Davier, Davis, Daw, Day, De, DeBra, Degallaix,
  De~Laurentis, Del\'eglise, Del~Pozzo, Demos, Denker, Dent, De~Pietri,
  Dergachev, De~Rosa, DeRosa, De~Rossi, DeSalvo, de~Varona, Devenson,
  Dhurandhar, D\'{\i}az, Dietrich, Di~Fiore, Di~Giovanni, Di~Girolamo,
  Di~Lieto, Di~Pace, Di~Palma, Di~Renzo, Doctor, Dolique, Donovan, Dooley,
  Doravari, Dorrington, Douglas, Dovale~\'Alvarez, Downes, Drago,
  Dreissigacker, Driggers, Du, Ducrot, Dudi, Dupej, Dwyer, Edo, Edwards,
  Effler, Eggenstein, Ehrens, Eichholz, Eikenberry, Eisenstein, Essick,
  Estevez, Etienne, Etzel, Evans, Evans, Factourovich, Fafone, Fair, Fairhurst,
  Fan, Farinon, Farr, Farr, Fauchon-Jones, Favata, Fays, Fee, Fehrmann, Feicht,
  Fejer, Fernandez-Galiana, Ferrante, Ferreira, Ferrini, Fidecaro, Finstad,
  Fiori, Fiorucci, Fishbach, Fisher, Fitz-Axen, Flaminio, Fletcher, Fong, Font,
  Forsyth, Forsyth, Fournier, Frasca, Frasconi, Frei, Freise, Frey, Frey,
  Fries, Fritschel, Frolov, Fulda, Fyffe, Gabbard, Gadre, Gaebel, Gair,
  Gammaitoni, Ganija, Gaonkar, Garcia-Quiros, Garufi, Gateley, Gaudio, Gaur,
  Gayathri, Gehrels, Gemme, Genin, Gennai, George, George, Gergely, Germain,
  Ghonge, Ghosh, Ghosh, Ghosh, Giaime, Giardina, Giazotto, Gill, Glover, Goetz,
  Goetz, Gomes, Goncharov, Gonz\'alez, Gonzalez~Castro, Gopakumar, Gorodetsky,
  Gossan, Gosselin, Gouaty, Grado, Graef, Granata, Grant, Gras, Gray, Greco,
  Green, Gretarsson, Groot, Grote, Grunewald, Gruning, Guidi, Guo, Gupta,
  Gupta, Gushwa, Gustafson, Gustafson, Halim, Hall, Hall, Hamilton, Hammond,
  Haney, Hanke, Hanks, Hanna, Hannam, Hannuksela, Hanson, Hardwick, Harms,
  Harry, Harry, Hart, Haster, Haughian, Healy, Heidmann, Heintze, Heitmann,
  Hello, Hemming, Hendry, Heng, Hennig, Heptonstall, Heurs, Hild, Hinderer, Ho,
  Hoak, Hofman, Holt, Holz, Hopkins, Horst, Hough, Houston, Howell, Hreibi, Hu,
  Huerta, Huet, Hughey, Husa, Huttner, Huynh-Dinh, Indik, Inta, Intini, Isa,
  Isac, Isi, Iyer, Izumi, Jacqmin, Jani, Jaranowski, Jawahar,
  Jim\'enez-Forteza, Johnson, Johnson-McDaniel, Jones, Jones, Jonker, Ju,
  Junker, Kalaghatgi, Kalogera, Kamai, Kandhasamy, Kang, Kanner, Kapadia,
  Karki, Karvinen, Kasprzack, Kastaun, Katolik, Katsavounidis, Katzman, Kaufer,
  Kawabe, K\'ef\'elian, Keitel, Kemball, Kennedy, Kent, Key, Khalili, Khan,
  Khan, Khan, Khazanov, Kijbunchoo, Kim, Kim, Kim, Kim, Kim, Kim, Kimbrell,
  King, King, Kinley-Hanlon, Kirchhoff, Kissel, Kleybolte, Klimenko, Knowles,
  Koch, Koehlenbeck, Koley, Kondrashov, Kontos, Korobko, Korth, Kowalska,
  Kozak, Kr\"amer, Kringel, Krishnan, Kr\'olak, Kuehn, Kumar, Kumar, Kumar,
  Kuo, Kutynia, Kwang, Lackey, Lai, Landry, Lang, Lange, Lantz, Lanza, Larson,
  Lartaux-Vollard, Lasky, Laxen, Lazzarini, Lazzaro, Leaci, Leavey, Lee, Lee,
  Lee, Lee, Lee, Lehmann, Lenon, Leon, Leonardi, Leroy, Letendre, Levin, Li,
  Linker, Littenberg, Liu, Liu, Lo, Lockerbie, London, Lord, Lorenzini,
  Loriette, Lormand, Losurdo, Lough, Lousto, Lovelace, L\"uck, Lumaca,
  Lundgren, Lynch, Ma, Macas, Macfoy, Machenschalk, MacInnis, Macleod, Maga\~na
  Hernandez, Maga\~na Sandoval, Maga\~na Zertuche, Magee, Majorana, Maksimovic,
  Man, Mandic, Mangano, Mansell, Manske, Mantovani, Marchesoni, Marion,
  M\'arka, M\'arka, Markakis, Markosyan, Markowitz, Maros, Marquina, Marsh,
  Martelli, Martellini, Martin, Martin, Martynov, Marx, Mason, Massera,
  Masserot, Massinger, Masso-Reid, Mastrogiovanni, Matas, Matichard, Matone,
  Mavalvala, Mazumder, McCarthy, McClelland, McCormick, McCuller, McGuire,
  McIntyre, McIver, McManus, McNeill, McRae, McWilliams, Meacher, Meadors,
  Mehmet, Meidam, Mejuto-Villa, Melatos, Mendell, Mercer, Merilh, Merzougui,
  Meshkov, Messenger, Messick, Metzdorff, Meyers, Miao, Michel, Middleton,
  Mikhailov, Milano, Miller, Miller, Miller, Millhouse, Milovich-Goff,
  Minazzoli, Minenkov, Ming, Mishra, Mitra, Mitrofanov, Mitselmakher,
  Mittleman, Moffa, Moggi, Mogushi, Mohan, Mohapatra, Molina, Montani, Moore,
  Moraru, Moreno, Morisaki, Morriss, Mours, Mow-Lowry, Mueller, Muir,
  Mukherjee, Mukherjee, Mukherjee, Mukund, Mullavey, Munch, Mu\~niz, Muratore,
  Murray, Nagar, Napier, Nardecchia, Naticchioni, Nayak, Neilson, Nelemans,
  Nelson, Nery, Neunzert, Nevin, Newport, Newton, Ng, Nguyen, Nguyen, Nichols,
  Nielsen, Nissanke, Nitz, Noack, Nocera, Nolting, North, Nuttall, Oberling,
  O'Dea, Ogin, Oh, Oh, Ohme, Okada, Oliver, Oppermann, Oram, O'Reilly,
  Ormiston, Ortega, O'Shaughnessy, Ossokine, Ottaway, Overmier, Owen, Pace,
  Page, Page, Pai, Pai, Palamos, Palashov, Palomba, Pal-Singh, Pan, Pan, Pang,
  Pang, Pankow, Pannarale, Pant, Paoletti, Paoli, Papa, Parida, Parker,
  Pascucci, Pasqualetti, Passaquieti, Passuello, Patil, Patricelli, Pearlstone,
  Pedraza, Pedurand, Pekowsky, Pele, Penn, Perez, Perreca, Perri, Pfeiffer,
  Phelps, Piccinni, Pichot, Piergiovanni, Pierro, Pillant, Pinard, Pinto,
  Pirello, Pitkin, Poe, Poggiani, Popolizio, Porter, Post, Powell, Prasad,
  Pratt, Pratten, Predoi, Prestegard, Prijatelj, Principe, Privitera, Prix,
  Prodi, Prokhorov, Puncken, Punturo, Puppo, P\"urrer, Qi, Quetschke, Quintero,
  Quitzow-James, Raab, Rabeling, Radkins, Raffai, Raja, Rajan, Rajbhandari,
  Rakhmanov, Ramirez, Ramos-Buades, Rapagnani, Raymond, Razzano, Read,
  Regimbau, Rei, Reid, Reitze, Ren, Reyes, Ricci, Ricker, Rieger, Riles, Rizzo,
  Robertson, Robie, Robinet, Rocchi, Rolland, Rollins, Roma, Romano, Romano,
  Romel, Romie, Rosi\ifmmode~\acute{n}\else \'{n}\fi{}ska, Ross, Rowan,
  R\"udiger, Ruggi, Rutins, Ryan, Sachdev, Sadecki, Sadeghian, Sakellariadou,
  Salconi, Saleem, Salemi, Samajdar, Sammut, Sampson, Sanchez, Sanchez,
  Sanchis-Gual, Sandberg, Sanders, Sassolas, Sathyaprakash, Saulson, Sauter,
  Savage, Sawadsky, Schale, Scheel, Scheuer, Schmidt, Schmidt, Schnabel,
  Schofield, Sch\"onbeck, Schreiber, Schuette, Schulte, Schutz, Schwalbe,
  Scott, Scott, Seidel, Sellers, Sengupta, Sentenac, Sequino, Sergeev,
  Shaddock, Shaffer, Shah, Shahriar, Shaner, Shao, Shapiro, Shawhan, Sheperd,
  Shoemaker, Shoemaker, Siellez, Siemens, Sieniawska, Sigg, Silva, Singer,
  Singh, Singhal, Sintes, Slagmolen, Smith, Smith, Smith, Somala, Son,
  Sonnenberg, Sorazu, Sorrentino, Souradeep, Spencer, Srivastava, Staats,
  Staley, Steinke, Steinlechner, Steinlechner, Steinmeyer, Stevenson, Stone,
  Stops, Strain, Stratta, Strigin, Strunk, Sturani, Stuver, Summerscales, Sun,
  Sunil, Suresh, Sutton, Swinkels, Szczepa\ifmmode~\acute{n}\else
  \'{n}\fi{}czyk, Tacca, Tait, Talbot, Talukder, Tanner, T\'apai, Taracchini,
  Tasson, Taylor, Taylor, Tewari, Theeg, Thies, Thomas, Thomas, Thomas, Thorne,
  Thorne, Thrane, Tiwari, Tiwari, Tokmakov, Toland, Tonelli, Tornasi,
  Torres-Forn\'e, Torrie, T\"oyr\"a, Travasso, Traylor, Trinastic, Tringali,
  Trozzo, Tsang, Tse, Tso, Tsukada, Tsuna, Tuyenbayev, Ueno, Ugolini,
  Unnikrishnan, Urban, Usman, Vahlbruch, Vajente, Valdes, Vallisneri, van
  Bakel, van Beuzekom, van~den Brand, Van Den~Broeck, Vander-Hyde, van~der
  Schaaf, van Heijningen, van Veggel, Vardaro, Varma, Vass, Vas\'uth, Vecchio,
  Vedovato, Veitch, Veitch, Venkateswara, Venugopalan, Verkindt, Vetrano,
  Vicer\'e, Viets, Vinciguerra, Vine, Vinet, Vitale, Vo, Vocca, Vorvick,
  Vyatchanin, Wade, Wade, Wade, Walet, Walker, Wallace, Walsh, Wang, Wang,
  Wang, Wang, Wang, Ward, Warner, Was, Watchi, Weaver, Wei, Weinert, Weinstein,
  Weiss, Wen, Wessel, We\ss{}els, Westerweck, Westphal, Wette, Whelan,
  Whitcomb, Whiting, Whittle, Wilken, Williams, Williams, Williamson, Willis,
  Willke, Wimmer, Winkler, Wipf, Wittel, Woan, Woehler, Wofford, Wong, Worden,
  Wright, Wu, Wysocki, Xiao, Yamamoto, Yancey, Yang, Yap, Yazback, Yu, Yu,
  Yvert, Zadro\ifmmode~\dot{z}\else \.{z}\fi{}ny, Zanolin, Zelenova, Zendri,
  Zevin, Zhang, Zhang, Zhang, Zhang, Zhao, Zhou, Zhou, Zhu, Zhu, Zimmerman,
  Zucker, \& Zweizig}]{PhysRevLett.119.161101}
Abbott, B.~P., Abbott, R., Abbott, T.~D., {et~al.} 2017, Phys. Rev. Lett., 119,
  161101

\bibitem[{Abbott {et~al.}(2018)}]{LIGOScientific:2018cki}
Abbott, B.~P. {et~al.} 2018, Phys. Rev. Lett., 121, 161101

\bibitem[{Agalianou \& Gourgouliatos(2023)}]{Agalianou:2023lvv}
Agalianou, V. \& Gourgouliatos, K.~N. 2023, \mnras, 522, 5879

\bibitem[{Aguilera {et~al.}(2004)Aguilera, Blaschke, \&
  Grigorian}]{Aguilera:2002dh}
Aguilera, D.~N., Blaschke, D., \& Grigorian, H. 2004, \aap, 416, 991

\bibitem[{{Alvarez-Castillo}(2021)}]{2021AN....342..234A}
{Alvarez-Castillo}, D.~E. 2021, Astronomische Nachrichten, 342, 234

\bibitem[{{Alvarez-Castillo} {et~al.}(2019){Alvarez-Castillo}, Antoniadis,
  Ayriyan, Blaschke, Danchev, Grigorian, Largani, \&
  Weber}]{Alvarez-Castillo:2019apz}
{Alvarez-Castillo}, D.~E., Antoniadis, J., Ayriyan, A., {et~al.} 2019,
  Astronomische Nachrichten, 340, 878

\bibitem[{Andersson {et~al.}(2005)Andersson, Glampedakis, Haskell, \&
  Watts}]{10.1111/j.1365-2966.2005.09167.x}
Andersson, N., Glampedakis, K., Haskell, B., \& Watts, A.~L. 2005, \mnras, 361,
  1153

\bibitem[{{Antoniadis}(2014)}]{antoniadis:apjl14}
{Antoniadis}, J. 2014, \apjl, 797, L24

\bibitem[{{Antoniadis}(2021)}]{antoniadis:2021}
{Antoniadis}, J. 2021, \mnras, 501, 1116

\bibitem[{{Antoniadis} {et~al.}(2013){Antoniadis}, {Freire}, {Wex}, {Tauris},
  {Lynch}, {van Kerkwijk}, {Kramer}, {Bassa}, {Dhillon}, {Driebe}, {Hessels},
  {Kaspi}, {Kondratiev}, {Langer}, {Marsh}, {McLaughlin}, {Pennucci}, {Ransom},
  {Stairs}, {van Leeuwen}, {Verbiest}, \& {Whelan}}]{antoniadis:2013sci}
{Antoniadis}, J., {Freire}, P. C.~C., {Wex}, N., {et~al.} 2013, Science, 340,
  448

\bibitem[{{Antoniadis} {et~al.}(2016){Antoniadis}, {Kaplan}, {Stovall},
  {Freire}, {Deneva}, {Koester}, {Jenet}, \& {Martinez}}]{Antoniadis:2016bnj}
{Antoniadis}, J., {Kaplan}, D.~L., {Stovall}, K., {et~al.} 2016, \apj, 830, 36

\bibitem[{{Astropy Collaboration} {et~al.}(2018){Astropy Collaboration},
  {Price-Whelan}, {Sip{\H{o}}cz}, {G{\"u}nther}, {Lim}, {Crawford}, {Conseil},
  {Shupe}, {Craig}, {Dencheva}, {Ginsburg}, {VanderPlas}, {Bradley},
  {P{\'e}rez-Su{\'a}rez}, {de Val-Borro}, {Aldcroft}, {Cruz}, {Robitaille},
  {Tollerud}, {Ardelean}, {Babej}, {Bach}, {Bachetti}, {Bakanov}, {Bamford},
  {Barentsen}, {Barmby}, {Baumbach}, {Berry}, {Biscani}, {Boquien}, {Bostroem},
  {Bouma}, {Brammer}, {Bray}, {Breytenbach}, {Buddelmeijer}, {Burke},
  {Calderone}, {Cano Rodr{\'\i}guez}, {Cara}, {Cardoso}, {Cheedella}, {Copin},
  {Corrales}, {Crichton}, {D'Avella}, {Deil}, {Depagne}, {Dietrich}, {Donath},
  {Droettboom}, {Earl}, {Erben}, {Fabbro}, {Ferreira}, {Finethy}, {Fox},
  {Garrison}, {Gibbons}, {Goldstein}, {Gommers}, {Greco}, {Greenfield},
  {Groener}, {Grollier}, {Hagen}, {Hirst}, {Homeier}, {Horton}, {Hosseinzadeh},
  {Hu}, {Hunkeler}, {Ivezi{\'c}}, {Jain}, {Jenness}, {Kanarek}, {Kendrew},
  {Kern}, {Kerzendorf}, {Khvalko}, {King}, {Kirkby}, {Kulkarni}, {Kumar},
  {Lee}, {Lenz}, {Littlefair}, {Ma}, {Macleod}, {Mastropietro}, {McCully},
  {Montagnac}, {Morris}, {Mueller}, {Mumford}, {Muna}, {Murphy}, {Nelson},
  {Nguyen}, {Ninan}, {N{\"o}the}, {Ogaz}, {Oh}, {Parejko}, {Parley}, {Pascual},
  {Patil}, {Patil}, {Plunkett}, {Prochaska}, {Rastogi}, {Reddy Janga},
  {Sabater}, {Sakurikar}, {Seifert}, {Sherbert}, {Sherwood-Taylor}, {Shih},
  {Sick}, {Silbiger}, {Singanamalla}, {Singer}, {Sladen}, {Sooley},
  {Sornarajah}, {Streicher}, {Teuben}, {Thomas}, {Tremblay}, {Turner},
  {Terr{\'o}n}, {van Kerkwijk}, {de la Vega}, {Watkins}, {Weaver}, {Whitmore},
  {Woillez}, {Zabalza}, \& {Astropy Contributors}}]{2018AJ....156..123A}
{Astropy Collaboration}, {Price-Whelan}, A.~M., {Sip{\H{o}}cz}, B.~M., {et~al.}
  2018, \aj, 156, 123

\bibitem[{{Astropy Collaboration} {et~al.}(2013){Astropy Collaboration},
  {Robitaille}, {Tollerud}, {Greenfield}, {Droettboom}, {Bray}, {Aldcroft},
  {Davis}, {Ginsburg}, {Price-Whelan}, {Kerzendorf}, {Conley}, {Crighton},
  {Barbary}, {Muna}, {Ferguson}, {Grollier}, {Parikh}, {Nair}, {Unther},
  {Deil}, {Woillez}, {Conseil}, {Kramer}, {Turner}, {Singer}, {Fox}, {Weaver},
  {Zabalza}, {Edwards}, {Azalee Bostroem}, {Burke}, {Casey}, {Crawford},
  {Dencheva}, {Ely}, {Jenness}, {Labrie}, {Lim}, {Pierfederici}, {Pontzen},
  {Ptak}, {Refsdal}, {Servillat}, \& {Streicher}}]{2013A&A...558A..33A}
{Astropy Collaboration}, {Robitaille}, T.~P., {Tollerud}, E.~J., {et~al.} 2013,
  \aap, 558, A33

\bibitem[{{Barr} {et~al.}(2013){Barr}, {Champion}, {Kramer}, {Eatough},
  {Freire}, {Karuppusamy}, {Lee}, {Verbiest}, {Bassa}, {Lyne}, {Stappers},
  {Lorimer}, \& {Klein}}]{Barr:mnras2013}
{Barr}, E.~D., {Champion}, D.~J., {Kramer}, M., {et~al.} 2013, \mnras, 435,
  2234

\bibitem[{Bauswein {et~al.}(2022)Bauswein, Blaschke, \&
  Fischer}]{Bauswein:2022vtq}
Bauswein, A., Blaschke, D., \& Fischer, T. 2022, Astrophysics in the XXI
  Century with Compact Stars (Singapore: World Scientific), 283–320

\bibitem[{{Baym} {et~al.}(1971){Baym}, {Pethick}, \&
  {Sutherland}}]{Baym:1971apj}
{Baym}, G., {Pethick}, C., \& {Sutherland}, P. 1971, \apj, 170, 299

\bibitem[{Bejger {et~al.}(2017)Bejger, Blaschke, Haensel, Zdunik, \&
  Fortin}]{Bejger:2016emu}
Bejger, M., Blaschke, D., Haensel, P., Zdunik, J.~L., \& Fortin, M. 2017, \aap,
  600, A39

\bibitem[{Benic {et~al.}(2015)Benic, Blaschke, Alvarez-Castillo, Fischer, \&
  Typel}]{Benic:2014jia}
Benic, S., Blaschke, D., Alvarez-Castillo, D.~E., Fischer, T., \& Typel, S.
  2015, \aap, 577, A40

\bibitem[{Berdermann {et~al.}(2006)Berdermann, Blaschke, Grigorian, \&
  Voskresensky}]{Berdermann:2006rk}
Berdermann, J., Blaschke, D., Grigorian, H., \& Voskresensky, D.~N. 2006, Prog.
  Part. Nucl. Phys., 57, 334

\bibitem[{{Bhattacharya} \& {van den Heuvel}(1991)}]{Bhattacharya:1991pre}
{Bhattacharya}, D. \& {van den Heuvel}, E.~P.~J. 1991, \physrep, 203, 1

\bibitem[{Bhattacharyya(2023)}]{Bhattacharyya:mdpi23}
Bhattacharyya, S. 2023, Galaxies, 11

\bibitem[{{Bhattacharyya} \& {Chakrabarty}(2017)}]{2017ApJ...835....4B}
{Bhattacharyya}, S. \& {Chakrabarty}, D. 2017, \apj, 835, 4

\bibitem[{{Bildsten} {et~al.}(1997){Bildsten}, {Chakrabarty}, {Chiu}, {Finger},
  {Koh}, {Nelson}, {Prince}, {Rubin}, {Scott}, {Stollberg}, {Vaughan},
  {Wilson}, \& {Wilson}}]{1997ApJS..113..367B}
{Bildsten}, L., {Chakrabarty}, D., {Chiu}, J., {et~al.} 1997, \apjs, 113, 367

\bibitem[{{Bill Wolf and Josiah Schwab}(2017)}]{pymesareader}
{Bill Wolf and Josiah Schwab}. 2017, Zenodo

\bibitem[{Bogdanov {et~al.}(2021)}]{Bogdanov:2021yip}
Bogdanov, S. {et~al.} 2021, \apjl, 914, L15

\bibitem[{Bombaci \& Popov(2004)}]{Bombaci:2004nu}
Bombaci, I. \& Popov, S.~B. 2004, \aap, 424, 627

\bibitem[{{Camilo} {et~al.}(2015){Camilo}, {Kerr}, {Ray}, {Ransom},
  {Sarkissian}, {Cromartie}, {Johnston}, {Reynolds}, {Wolff}, {Freire},
  {Bhattacharyya}, {Ferrara}, {Keith}, {Michelson}, {Saz Parkinson}, \&
  {Wood}}]{Camilo:apj2015}
{Camilo}, F., {Kerr}, M., {Ray}, P.~S., {et~al.} 2015, \apj, 810, 85

\bibitem[{Carlomagno {et~al.}(2024{\natexlab{a}})Carlomagno, Contrera,
  Grunfeld, \& Blaschke}]{Carlomagno:2024vvr}
Carlomagno, J.~P., Contrera, G.~A., Grunfeld, A.~G., \& Blaschke, D.
  2024{\natexlab{a}}, Universe, 10, 336

\bibitem[{Carlomagno {et~al.}(2024{\natexlab{b}})Carlomagno, Contrera,
  Grunfeld, \& Blaschke}]{Carlomagno:2023nrc}
Carlomagno, J.~P., Contrera, G.~A., Grunfeld, A.~G., \& Blaschke, D.
  2024{\natexlab{b}}, Phys. Rev. D, 109, 043050

\bibitem[{{Chaboyer} \& {Zahn}(1992)}]{1992A&A...253..173C}
{Chaboyer}, B. \& {Zahn}, J.~P. 1992, \aap, 253, 173

\bibitem[{{Champion} {et~al.}(2008){Champion}, {Ransom}, {Lazarus}, {Camilo},
  {Bassa}, {Kaspi}, {Nice}, {Freire}, {Stairs}, {van Leeuwen}, {Stappers},
  {Cordes}, {Hessels}, {Lorimer}, {Arzoumanian}, {Backer}, {Bhat},
  {Chatterjee}, {Cognard}, {Deneva}, {Faucher-Gigu{\`e}re}, {Gaensler}, {Han},
  {Jenet}, {Kasian}, {Kondratiev}, {Kramer}, {Lazio}, {McLaughlin},
  {Venkataraman}, \& {Vlemmings}}]{Champion:sci2008}
{Champion}, D.~J., {Ransom}, S.~M., {Lazarus}, P., {et~al.} 2008, Science, 320,
  1309

\bibitem[{Chandra(2025)}]{2023arXiv231113303D}
Chandra, A.~D. 2025, Publications of the Astronomical Society of Australia
  (accepted) [\eprint[arXiv]{2311.13303}]

\bibitem[{{Chevalier}(1989)}]{1989ApJ...346..847C}
{Chevalier}, R.~A. 1989, \apj, 346, 847

\bibitem[{{Corbet}(1996)}]{1996ApJ...457L..31C}
{Corbet}, R. H.~D. 1996, \apjl, 457, L31

\bibitem[{{Cromartie} {et~al.}(2020){Cromartie}, {Fonseca}, {Ransom},
  {Demorest}, {Arzoumanian}, {Blumer}, {Brook}, {DeCesar}, {Dolch}, {Ellis},
  {Ferdman}, {Ferrara}, {Garver-Daniels}, {Gentile}, {Jones}, {Lam}, {Lorimer},
  {Lynch}, {McLaughlin}, {Ng}, {Nice}, {Pennucci}, {Spiewak}, {Stairs},
  {Stovall}, {Swiggum}, \& {Zhu}}]{2020NatAs...4...72C}
{Cromartie}, H.~T., {Fonseca}, E., {Ransom}, S.~M., {et~al.} 2020, Nature
  Astronomy, 4, 72

\bibitem[{{Cumming} {et~al.}(2001){Cumming}, {Zweibel}, \&
  {Bildsten}}]{2001ApJ...557..958C}
{Cumming}, A., {Zweibel}, E., \& {Bildsten}, L. 2001, \apj, 557, 958

\bibitem[{{Deneva} {et~al.}(2013){Deneva}, {Stovall}, {McLaughlin}, {Bates},
  {Freire}, {Martinez}, {Jenet}, \& {Bagchi}}]{Deneva:apj2013}
{Deneva}, J.~S., {Stovall}, K., {McLaughlin}, M.~A., {et~al.} 2013, \apj, 775,
  51

\bibitem[{{Elsner} {et~al.}(1980){Elsner}, {Ghosh}, \&
  {Lamb}}]{1980ApJ...241L.155E}
{Elsner}, R.~F., {Ghosh}, P., \& {Lamb}, F.~K. 1980, \apjl, 241, L155

\bibitem[{{Freire} {et~al.}(2011){Freire}, {Bassa}, {Wex}, {Stairs},
  {Champion}, {Ransom}, {Lazarus}, {Kaspi}, {Hessels}, {Kramer}, {Cordes},
  {Verbiest}, {Podsiadlowski}, {Nice}, {Deneva}, {Lorimer}, {Stappers},
  {McLaughlin}, \& {Camilo}}]{2011MNRAS.412.2763F}
{Freire}, P.~C.~C., {Bassa}, C.~G., {Wex}, N., {et~al.} 2011, \mnras, 412, 2763

\bibitem[{{Freire} \& {Tauris}(2014)}]{Freire:mnras14}
{Freire}, P. C.~C. \& {Tauris}, T.~M. 2014, \mnras, 438, L86

\bibitem[{Fukushima \& Yu(2024)}]{Fukushima:2024cpg}
Fukushima, K. \& Yu, C. 2024 [\eprint[arXiv]{2401.04568}]

\bibitem[{{Galloway} {et~al.}(2002){Galloway}, {Chakrabarty}, {Morgan}, \&
  {Remillard}}]{2002ApJ...576L.137G}
{Galloway}, D.~K., {Chakrabarty}, D., {Morgan}, E.~H., \& {Remillard}, R.~A.
  2002, \apjl, 576, L137

\bibitem[{Garrett(2021)}]{SciencePlots}
Garrett, J.~D. 2021

\bibitem[{Gerlach(1968)}]{Gerlach:1968zz}
Gerlach, U.~H. 1968, Phys. Rev., 172, 1325

\bibitem[{{Ginzburg} \& {Chiang}(2022)}]{ginzburg:mnras21}
{Ginzburg}, S. \& {Chiang}, E. 2022, \mnras, 509, L1

\bibitem[{{Glebbeek} {et~al.}(2009){Glebbeek}, {Gaburov}, {de Mink}, {Pols}, \&
  {Portegies Zwart}}]{Glebbeek:aap2009}
{Glebbeek}, E., {Gaburov}, E., {de Mink}, S.~E., {Pols}, O.~R., \& {Portegies
  Zwart}, S.~F. 2009, \aap, 497, 255

\bibitem[{Glendenning {et~al.}(1997)Glendenning, Pei, \&
  Weber}]{Glendenning:1997fy}
Glendenning, N.~K., Pei, S., \& Weber, F. 1997, Phys. Rev. Lett., 79, 1603

\bibitem[{Harris {et~al.}(2020)Harris, Millman, van~der Walt, Gommers,
  Virtanen, Cournapeau, Wieser, Taylor, Berg, Smith, Kern, Picus, Hoyer, van
  Kerkwijk, Brett, Haldane, del R{\'{i}}o, Wiebe, Peterson,
  G{\'{e}}rard-Marchant, Sheppard, Reddy, Weckesser, Abbasi, Gohlke, \&
  Oliphant}]{harris2020array:numpy}
Harris, C.~R., Millman, K.~J., van~der Walt, S.~J., {et~al.} 2020, Nature, 585,
  357

\bibitem[{{Harrison} \& {Tademaru}(1975)}]{1975ApJ...201..447H}
{Harrison}, E.~R. \& {Tademaru}, E. 1975, \apj, 201, 447

\bibitem[{{Hebeler} {et~al.}(2013){Hebeler}, {Lattimer}, {Pethick}, \&
  {Schwenk}}]{Hebeler:2013apj}
{Hebeler}, K., {Lattimer}, J.~M., {Pethick}, C.~J., \& {Schwenk}, A. 2013,
  \apj, 773, 11

\bibitem[{{Heger} {et~al.}(2000){Heger}, {Langer}, \&
  {Woosley}}]{Heger:apj2000}
{Heger}, A., {Langer}, N., \& {Woosley}, S.~E. 2000, \apj, 528, 368

\bibitem[{{Heger} {et~al.}(2005){Heger}, {Woosley}, \&
  {Spruit}}]{Heger:apj2005}
{Heger}, A., {Woosley}, S.~E., \& {Spruit}, H.~C. 2005, \apj, 626, 350

\bibitem[{Hempel {et~al.}(2016)Hempel, Heinimann, Yudin, Iosilevskiy,
  Liebend\"orfer, \& Thielemann}]{Hempel:2015vlg}
Hempel, M., Heinimann, O., Yudin, A., {et~al.} 2016, Phys. Rev. D, 94, 103001

\bibitem[{{Henyey} {et~al.}(1965){Henyey}, {Vardya}, \&
  {Bodenheimer}}]{Henyey:apj1965}
{Henyey}, L., {Vardya}, M.~S., \& {Bodenheimer}, P. 1965, \apj, 142, 841

\bibitem[{{Hills}(1983)}]{1983apj:Hills}
{Hills}, J.~G. 1983, \apj, 267, 322

\bibitem[{Hobbs {et~al.}(2005)Hobbs, Lorimer, Lyne, \& Kramer}]{Hobbs:2005yx}
Hobbs, G., Lorimer, D.~R., Lyne, A.~G., \& Kramer, M. 2005, \mnras, 360, 974

\bibitem[{Hunter(2007)}]{Hunter:2007:matplotlib}
Hunter, J.~D. 2007, Computing in Science \& Engineering, 9, 90

\bibitem[{{Iglesias} \& {Rogers}(1996)}]{opal_opacities}
{Iglesias}, C.~A. \& {Rogers}, F.~J. 1996, \apj, 464, 943

\bibitem[{{Jiang} {et~al.}(2015){Jiang}, {Li}, {Dey}, \& {Dey}}]{Jiang:apj15}
{Jiang}, L., {Li}, X.-D., {Dey}, J., \& {Dey}, M. 2015, \apj, 807, 41

\bibitem[{{Jiang} {et~al.}(2020){Jiang}, {Wang}, {Chen}, {Li}, {Liu}, \&
  {Gao}}]{2020A&A...633A..45J}
{Jiang}, L., {Wang}, N., {Chen}, W.-C., {et~al.} 2020, \aap, 633, A45

\bibitem[{{Jiang} {et~al.}(2021){Jiang}, {Wang}, {Chen}, {Liu}, {Leng}, {Yuan},
  \& {Qian}}]{Jiang:raa2021}
{Jiang}, L., {Wang}, N., {Chen}, W.-C., {et~al.} 2021, Research in Astronomy
  and Astrophysics, 21, 231

\bibitem[{Kalapotharakos {et~al.}(2021)Kalapotharakos, Wadiasingh, Harding, \&
  Kazanas}]{Kalapotharakos:2020rmz}
Kalapotharakos, C., Wadiasingh, Z., Harding, A.~K., \& Kazanas, D. 2021, \apj,
  907, 63

\bibitem[{Kaminski {et~al.}(2016)Kaminski, Uhlemann, Bleicher, \&
  Schaffner-Bielich}]{Kaminski:2014jda}
Kaminski, M., Uhlemann, C.~F., Bleicher, M., \& Schaffner-Bielich, J. 2016,
  Phys. Lett. B, 760, 170

\bibitem[{Khosravi~Largani {et~al.}(2024)Khosravi~Largani, Fischer, Shibagaki,
  Cerdá-Durán, \& Torres-Forné}]{2023arXiv231115992K}
Khosravi~Largani, N., Fischer, T., Shibagaki, S., Cerdá-Durán, P., \&
  Torres-Forné, A. 2024, \aap, 687, A245

\bibitem[{Kluyver {et~al.}(2016)Kluyver, Ragan-Kelley, P{\'e}rez, Granger,
  Bussonnier, Frederic, Kelley, Hamrick, Grout, Corlay, Ivanov, Avila, Abdalla,
  \& Willing}]{Kluyver2016jupyter}
Kluyver, T., Ragan-Kelley, B., P{\'e}rez, F., {et~al.} 2016, in Positioning and
  Power in Academic Publishing: Players, Agents and Agendas, ed. F.~Loizides \&
  B.~Schmidt, IOS Press, 87 -- 90

\bibitem[{{Knispel} {et~al.}(2015){Knispel}, {Lyne}, {Stappers}, {Freire},
  {Lazarus}, {Allen}, {Aulbert}, {Bock}, {Bogdanov}, {Brazier}, {Camilo},
  {Cardoso}, {Chatterjee}, {Cordes}, {Crawford}, {Deneva}, {Eggenstein},
  {Fehrmann}, {Ferdman}, {Hessels}, {Jenet}, {Karako-Argaman}, {Kaspi}, {van
  Leeuwen}, {Lorimer}, {Lynch}, {Machenschalk}, {Madsen}, {McLaughlin},
  {Patel}, {Ransom}, {Scholz}, {Siemens}, {Spitler}, {Stairs}, {Stovall},
  {Swiggum}, {Venkataraman}, {Wharton}, \& {Zhu}}]{Knispel:apj2015}
{Knispel}, B., {Lyne}, A.~G., {Stappers}, B.~W., {et~al.} 2015, \apj, 806, 140

\bibitem[{{Kolb} \& {Ritter}(1990)}]{Kolb:aap1990}
{Kolb}, U. \& {Ritter}, H. 1990, \aap, 236, 385

\bibitem[{Konar \& Choudhuri(2004)}]{10.1111/j.1365-2966.2004.07397.x}
Konar, S. \& Choudhuri, A.~R. 2004, \mnras, 348, 661

\bibitem[{{Langer}(1991)}]{Langer:aap1991}
{Langer}, N. 1991, \aap, 252, 669

\bibitem[{{Laplace}(2022)}]{2022A&C....3800516L}
{Laplace}, E. 2022, Astronomy and Computing, 38, 100516

\bibitem[{{Ledoux}(1947)}]{Ledoux:apj1947}
{Ledoux}, P. 1947, \apj, 105, 305

\bibitem[{Manchester {et~al.}(2005)Manchester, Hobbs, Teoh, \&
  Hobbs}]{Manchester_2005}
Manchester, R.~N., Hobbs, G.~B., Teoh, A., \& Hobbs, M. 2005, \aj, 129, 1993

\bibitem[{Miller {et~al.}(2019)}]{Miller:2019cac}
Miller, M.~C. {et~al.} 2019, \apjl, 887, L24

\bibitem[{Miller {et~al.}(2021)}]{Miller:2021qha}
Miller, M.~C. {et~al.} 2021, \apjl, 918, L28

\bibitem[{Mishustin {et~al.}(2003)Mishustin, Hanauske, Bhattacharyya, Satarov,
  Stoecker, \& Greiner}]{Mishustin:2002xe}
Mishustin, I.~N., Hanauske, M., Bhattacharyya, A., {et~al.} 2003, Phys. Lett.
  B, 552, 1

\bibitem[{{Misra} {et~al.}(2020){Misra}, {Fragos}, {Tauris}, {Zapartas}, \&
  {Aguilera-Dena}}]{2020A&A...642A.174M}
{Misra}, D., {Fragos}, T., {Tauris}, T.~M., {Zapartas}, E., \& {Aguilera-Dena},
  D.~R. 2020, \aap, 642, A174

\bibitem[{{Nugis} \& {Lamers}(2000)}]{Nugis:aap2000}
{Nugis}, T. \& {Lamers}, H.~J.~G.~L.~M. 2000, \aap, 360, 227

\bibitem[{{Nurmamat} {et~al.}(2019){Nurmamat}, {Zhu}, {L{\"u}}, {Wang}, {Li},
  \& {Liu}}]{2019JApA...40...32N}
{Nurmamat}, N., {Zhu}, C., {L{\"u}}, G., {et~al.} 2019, Journal of Astrophysics
  and Astronomy, 40, 32

\bibitem[{{Octau} {et~al.}(2018){Octau}, {Cognard}, {Guillemot}, {Tauris},
  {Freire}, {Desvignes}, \& {Theureau}}]{Octau:aap2018}
{Octau}, F., {Cognard}, I., {Guillemot}, L., {et~al.} 2018, \aap, 612, A78

\bibitem[{pandas~development team(2020)}]{reback2020pandas}
pandas~development team, T. 2020, pandas-dev/pandas: Pandas

\bibitem[{{Papitto} \& {de Martino}(2022)}]{Papitto:2022}
{Papitto}, A. \& {de Martino}, D. 2022, in Astrophysics and Space Science
  Library, Vol. 465, Astrophysics and Space Science Library, ed.
  S.~{Bhattacharyya}, A.~{Papitto}, \& D.~{Bhattacharya}, 157--200

\bibitem[{{Paschalidis} {et~al.}(2018){Paschalidis}, {Yagi},
  {Alvarez-Castillo}, {Blaschke}, \& {Sedrakian}}]{Paschalidis:2018prd}
{Paschalidis}, V., {Yagi}, K., {Alvarez-Castillo}, D., {Blaschke}, D.~B., \&
  {Sedrakian}, A. 2018, \prd, 97, 084038

\bibitem[{Paxton {et~al.}(2011)Paxton, Bildsten, Dotter, Herwig, Lesaffre, \&
  Timmes}]{Paxton:2010ji}
Paxton, B., Bildsten, L., Dotter, A., {et~al.} 2011, \apjs, 192, 3

\bibitem[{Paxton {et~al.}(2013)Paxton, Cantiello, Arras, Bildsten, Brown,
  Dotter, Mankovich, Montgomery, Stello, Timmes, \& Townsend}]{Paxton:2013pj}
Paxton, B., Cantiello, M., Arras, P., {et~al.} 2013, \apjs, 208, 4

\bibitem[{Paxton {et~al.}(2015)Paxton, Marchant, Schwab, Bauer, Bildsten,
  Cantiello, Dessart, Farmer, Hu, Langer, Townsend, Townsley, \&
  Timmes}]{Paxton:2015jva}
Paxton, B., Marchant, P., Schwab, J., {et~al.} 2015, \apjs, 220, 15

\bibitem[{Paxton {et~al.}(2018)Paxton, Schwab, Bauer, Bildsten, Blinnikov,
  Duffell, Farmer, Goldberg, Marchant, Sorokina, Thoul, Townsend, \&
  Timmes}]{Paxton:2017eie}
Paxton, B., Schwab, J., Bauer, E.~B., {et~al.} 2018, \apjs, 234, 34

\bibitem[{{Phinney}(1992)}]{Phinney:1992}
{Phinney}, E.~S. 1992, Philosophical Transactions of the Royal Society of
  London Series A, 341, 39

\bibitem[{{Portegies Zwart} {et~al.}(2011){Portegies Zwart}, {van den Heuvel},
  {van Leeuwen}, \& {Nelemans}}]{2011ApJ...734...55P}
{Portegies Zwart}, S., {van den Heuvel}, E.~P.~J., {van Leeuwen}, J., \&
  {Nelemans}, G. 2011, \apj, 734, 55

\bibitem[{Pradhan {et~al.}(2024)Pradhan, Chatterjee, \&
  Alvarez-Castillo}]{2023arXiv230908775K}
Pradhan, B.~K., Chatterjee, D., \& Alvarez-Castillo, D.~E. 2024, Monthly
  Notices of the Royal Astronomical Society, 531, 4640–4655

\bibitem[{{Rappaport} {et~al.}(1983){Rappaport}, {Verbunt}, \&
  {Joss}}]{Rappaport:apj1983}
{Rappaport}, S., {Verbunt}, F., \& {Joss}, P.~C. 1983, \apj, 275, 713

\bibitem[{{Rappaport} {et~al.}(2004){Rappaport}, {Fregeau}, \&
  {Spruit}}]{2004ApJ...606..436R}
{Rappaport}, S.~A., {Fregeau}, J.~M., \& {Spruit}, H. 2004, \apj, 606, 436

\bibitem[{{Reimers}(1975)}]{Reimers:bk1975}
{Reimers}, D. 1975, in Problems in stellar atmospheres and envelopes., 229--256

\bibitem[{Riley {et~al.}(2019)}]{Riley:2019yda}
Riley, T.~E. {et~al.} 2019, \apjl, 887, L21

\bibitem[{Sandin \& Blaschke(2007)}]{Sandin:2007zr}
Sandin, F. \& Blaschke, D. 2007, Phys. Rev. D, 75, 125013

\bibitem[{Schertler {et~al.}(2000)Schertler, Greiner, Schaffner-Bielich, \&
  Thoma}]{Schertler:2000xq}
Schertler, K., Greiner, C., Schaffner-Bielich, J., \& Thoma, M.~H. 2000, Nucl.
  Phys. A, 677, 463

\bibitem[{{Shibazaki} {et~al.}(1989){Shibazaki}, {Murakami}, {Shaham}, \&
  {Nomoto}}]{1989Natur.342..656S}
{Shibazaki}, N., {Murakami}, T., {Shaham}, J., \& {Nomoto}, K. 1989, Nature,
  342, 656

\bibitem[{{Spruit}(2002)}]{Spruit:aap2002}
{Spruit}, H.~C. 2002, \aap, 381, 923

\bibitem[{{Stovall} {et~al.}(2019){Stovall}, {Freire}, {Antoniadis}, {Bagchi},
  {Deneva}, {Garver-Daniels}, {Martinez}, {McLaughlin}, {Arzoumanian},
  {Blumer}, {Brook}, {Cromartie}, {Demorest}, {DeCesar}, {Dolch}, {Ellis},
  {Ferdman}, {Ferrara}, {Fonseca}, {Gentile}, {Jones}, {Lam}, {Lorimer},
  {Lynch}, {Ng}, {Nice}, {Pennucci}, {Ransom}, {Spiewak}, {Stairs}, {Swiggum},
  {Vigeland}, \& {Zhu}}]{Stovall:apj2019}
{Stovall}, K., {Freire}, P.~C.~C., {Antoniadis}, J., {et~al.} 2019, \apj, 870,
  74

\bibitem[{{Strader} {et~al.}(2019){Strader}, {Swihart}, {Chomiuk}, {Bahramian},
  {Britt}, {Cheung}, {Dage}, {Halpern}, {Li}, {Mignani}, {Orosz}, {Peacock},
  {Salinas}, {Shishkovsky}, \& {Tremou}}]{Strader:2019}
{Strader}, J., {Swihart}, S., {Chomiuk}, L., {et~al.} 2019, \apj, 872, 42

\bibitem[{{Sun} {et~al.}(2019){Sun}, {Li}, {Liu}, {L{\"u}}, {Wang}, \&
  {Zhu}}]{2019PASA...36....5S}
{Sun}, S., {Li}, L., {Liu}, H., {et~al.} 2019, \pasa, 36, e005

\bibitem[{Tauris(2012)}]{Tauris:sc2012}
Tauris, T.~M. 2012, Science, 335, 561

\bibitem[{{Tauris} {et~al.}(2017){Tauris}, {Kramer}, {Freire}, {Wex}, {Janka},
  {Langer}, {Podsiadlowski}, {Bozzo}, {Chaty}, {Kruckow}, {van den Heuvel},
  {Antoniadis}, {Breton}, \& {Champion}}]{Tauris:2017apj}
{Tauris}, T.~M., {Kramer}, M., {Freire}, P.~C.~C., {et~al.} 2017, \apj, 846,
  170

\bibitem[{{Tauris} \& {Savonije}(1999)}]{Tauris:aap1999}
{Tauris}, T.~M. \& {Savonije}, G.~J. 1999, \aap, 350, 928

\bibitem[{{Tauris} \& {van den Heuvel}(2006)}]{Tauris:bk2006}
{Tauris}, T.~M. \& {van den Heuvel}, E.~P.~J. 2006, in Compact stellar X-ray
  sources, Vol.~39, 623--665

\bibitem[{Tauris \& {van den Heuvel}(2023)}]{Tauris:2023nmj}
Tauris, T.~M. \& {van den Heuvel}, E. P.~J. 2023, Physics of {{Binary Star
  Evolution}}. {{From Stars}} to {{X-ray Binaries}} and {{Gravitational Wave
  Sources}}

\bibitem[{{Verbunt} \& {Phinney}(1995)}]{Verbunt:aa1995}
{Verbunt}, F. \& {Phinney}, E.~S. 1995, \aap, 296, 709

\bibitem[{{Verbunt} {et~al.}(1987){Verbunt}, {van den Heuvel}, {van Paradijs},
  \& {Rappaport}}]{1987Natur.329..312V}
{Verbunt}, F., {van den Heuvel}, E.~P.~J., {van Paradijs}, J., \& {Rappaport},
  S.~A. 1987, Nature, 329, 312

\bibitem[{Virtanen {et~al.}(2020)Virtanen, Gommers, Oliphant, Haberland, Reddy,
  Cournapeau, Burovski, Peterson, Weckesser, Bright, {van der Walt}, Brett,
  Wilson, Millman, Mayorov, Nelson, Jones, Kern, Larson, Carey, Polat, Feng,
  Moore, {VanderPlas}, Laxalde, Perktold, Cimrman, Henriksen, Quintero, Harris,
  Archibald, Ribeiro, Pedregosa, {van Mulbregt}, \& {SciPy 1.0
  Contributors}}]{2020SciPy-NMeth}
Virtanen, P., Gommers, R., Oliphant, T.~E., {et~al.} 2020, Nature Methods, 17,
  261

\bibitem[{{Watts} {et~al.}(2015){Watts}, {Espinoza}, {Xu}, {Andersson},
  {Antoniadis}, {Antonopoulou}, {Buchner}, {Datta}, {Demorest}, {Freire},
  {Hessels}, {Margueron}, {Oertel}, {Patruno}, {Possenti}, {Ransom}, {Stairs},
  \& {Stappers}}]{watts:2015}
{Watts}, A., {Espinoza}, C.~M., {Xu}, R., {et~al.} 2015, in Advancing
  Astrophysics with the Square Kilometre Array (AASKA14), 43

\bibitem[{{Zahn}(1977)}]{1977A&A....57..383Z}
{Zahn}, J.~P. 1977, \aap, 57, 383

\bibitem[{Zdunik {et~al.}(2006)Zdunik, Bejger, Haensel, \&
  Gourgoulhon}]{Zdunik:2005kh}
Zdunik, J.~L., Bejger, M., Haensel, P., \& Gourgoulhon, E. 2006, Astron.
  Astrophys., 450, 747

\end{thebibliography}


\end{document}